%% file: beyond_the_sum.tex
\documentclass{article}

\usepackage{arxiv}
\usepackage[utf8]{inputenc} 
\usepackage[T1]{fontenc}    
\usepackage{hyperref}       
\usepackage{url}            
\usepackage{booktabs}       
\usepackage{amsfonts}       
\usepackage{nicefrac}       
\usepackage{microtype}      
\usepackage{cleveref}       
\usepackage{lipsum}         
\usepackage{graphicx}
\usepackage{natbib}
\bibpunct{[}{]}{,}{n}{}{} 
\setcitestyle{numbers,square}
\usepackage{doi}

\usepackage{tikz}
\usepackage[edges]{forest} 
\usepackage{adjustbox}

\definecolor{framework-blue}{RGB}{0,114,178}

\forestset{
    mindmap/.style={
        for tree={
            grow=east,
            grow'=0,
            forked edge,
            draw,
            rounded corners,
            node options={align=center},
            text width=2.7cm,
            s sep=6pt,
            l sep=1.2cm,
            fork sep=1cm,
            child anchor=west,
            parent anchor=east,
            anchor=west,
            edge path={
                \noexpand\path[\forestoption{edge}]
                (!u.east) -- +(3pt,0) |- (.west)\forestoption{edge label};
            },
        }
    }
}

\tikzset{%
    parent/.style = {align=center,text width=2.5cm,rounded corners=3pt, line width=0.3mm, fill=gray!10,draw=gray!80},
    node/.style = {align=center,text width=2.2cm,rounded corners=3pt, fill=white,draw=framework-blue,line width=0.3mm},   
    leaf/.style = {align=center, text width=4.5cm,rounded corners=3pt, fill=white,draw=framework-blue,line width=0.3mm},
}

\title{Beyond the Sum: Unlocking AI Agents Potential Through Market Forces}

\author{
    \textbf{Jordi Montes Sanabria} \\
    Fewsats \\
    \texttt{jordi@fewsats.com} \and
    \textbf{Pol Alvarez Vecino} \\
    Fewsats \\
    \texttt{pol@fewsats.com}
}

\hypersetup{
    pdftitle={Beyond the Sum: Unlocking AI Agents Potential Through Market Forces},
    pdfsubject={Artificial Intelligence},
    pdfauthor={Jordi Montes Sanabria and Pol Alvarez Vecino},
    pdfkeywords={AI agents, Digital Infrastructure, Economic Systems, Emergent Intelligence, Market Dynamics},
    colorlinks=true,
    linkcolor=framework-blue,
    filecolor=framework-blue,
    urlcolor=framework-blue,
    citecolor=framework-blue 
}

\begin{document}
\maketitle



\begin{abstract}

The emergence of Large Language Models has fundamentally transformed the capabilities of AI agents, enabling a new class of autonomous agents capable of interacting with their environment through dynamic code generation and execution. These agents possess the theoretical capacity to operate as independent economic actors within digital markets, offering unprecedented potential for value creation through their distinct advantages in operational continuity, perfect replication, and distributed learning capabilities. However, contemporary digital infrastructure, architected primarily for human interaction, presents significant barriers to their participation.

This work presents a systematic analysis of the infrastructure requirements necessary for AI agents to function as autonomous participants in digital markets. We examine four key areas - identity and authorization, service discovery, interfaces, and payment systems - to show how existing infrastructure actively impedes agent participation. We argue that addressing these infrastructure challenges represents more than a technical imperative; it constitutes a fundamental step toward enabling new forms of economic organization. Much as traditional markets enable human intelligence to coordinate complex activities beyond individual capability, markets incorporating AI agents could dramatically enhance economic efficiency through continuous operation, perfect information sharing, and rapid adaptation to changing conditions. The infrastructure challenges identified in this work represent key barriers to realizing this potential.
\end{abstract}



\section{Introduction}

The field of AI agents has evolved significantly over decades, from early symbolic systems to reactive agents and reinforcement learning approaches \cite{Bengio1995,Botvinick2017}. Deep learning dramatically enhanced reinforcement learning capabilities, leading to breakthroughs like AlphaGo Zero \cite{Clark2014, Mnih2013, Silver2017}. This system mastered Go through pure self-play, without human examples \cite{Silver2017}. Yet these systems faced fundamental limitations. They required extensive training for each new task \cite{Marcus2018, Bergstra2012} and struggled to transfer knowledge between domains \cite{Goodfellow2013,Marcus2018}.

The current iteration of AI agents is powered by large language models. These models serve as the "brain" of modern AI agents, providing reasoning and decision-making capabilities that guide the actions of the agent. Unlike traditional AI approaches that require specific training for each task, LLM-based agents can understand and adapt to new situations through their broad knowledge of language and concepts \cite{Vaswani2017,Brown2020}. This flexibility stems from their exposure to diverse human knowledge during training, enabling them to reason about problems in ways similar to human thinking \cite{Kaplan2020,Ziegler2019}.

To interact with their environment, these agents use specialized components for perception and action. On the perception side, they can process diverse inputs—like images, text, or structured data—which are then provided to the LLM in a format it can understand, usually text \cite{Baltrušaitis2017}. On the output side, the LLM generates text that guides the agent's actions, whether through code generation to interact with APIs and digital systems, or through specific commands to use tools and manipulate the environment \cite{Ahuja2019}. This combination of flexible reasoning with the ability to perceive and act enables these agents to tackle complex, open-ended tasks \cite{Nyatsanga2023, Lee2022}.

Minecraft, a popular sandbox video game, presents players with an open-ended world where they must explore, gather resources, craft tools, and build complex structures to survive and thrive. The game's combination of simple rules and unlimited possibilities makes it an ideal testing ground for autonomous agents \cite{Kanervisto2020a,Kanervisto2020b,Kanervisto2020c}. The breakthrough system Voyager demonstrates the potential of LLM-based agents in this challenging environment. Through continuous exploration and skill development, it achieves true autonomous behavior—discovering new possibilities, writing code to implement solutions, and building upon its capabilities over time without human intervention \cite{Milani2023,Karttunen2019}.

There are many different approaches for structuring these AI agents. For instance, Voyager \cite{voyager} employs curriculum learning to progressively enhance its capabilities. Some agents output code designed to call specific functions \cite{claudette2024toolloop}, while others generate executable code in languages like Python or Bash\cite{voyager}. Input modalities also vary, with some agents processing only text descriptions\cite{zhu2023ghost}, while others are multimodal, integrating visual and textual inputs \cite{replit2024agent}. Despite these differences, these systems can be understood through a common framework of three essential parts \cite{ai_agents_survey}: the brain (where the LLM acts as a central controller), the perception components (which transform environmental information into inputs the LLM can understand), and the action components (which translate the LLM's outputs into concrete actions). This perception-reasoning-action framework is specially useful

In the same way as these agents can operate autonomously in game environments, they can potentially act as independent participants in the digital economy. The critical enabler is their code generation capability—they can dynamically create and execute programs to interact with digital services through APIs and automation tools. This ability to programmatically interface with services mirrors how modern digital economies operate, where both human developers and businesses increasingly rely on APIs and automated workflows to participate in markets. Just as Voyager autonomously discovers resources, crafts tools, and builds new capabilities in Minecraft's economy, an AI agent could generate code to discover services, execute transactions, or create new digital assets in the broader digital economy.

The promise of AI agents in digital markets is compelling: they could enable new forms of automated value creation, facilitate complex market interactions at machine speed, and unlock novel economic opportunities through their ability to rapidly adapt and innovate. However, even as AI agents gain these capabilities, they face significant infrastructure barriers. Current digital systems were designed with human operators in mind, embedding assumptions about human-scale reaction times, human-readable interfaces, and human-centric security models. These limitations constrain individual agents and add significant friction to their operation, creating a fundamental barrier to realizing this economic potential. Addressing these infrastructure challenges early is crucial to enable the emergence of truly efficient digital markets where AI agents can participate fully.

These infrastructure challenges represent a critical bottleneck in realizing the full potential of AI agents as economic participants.

The contribution of this paper is to examine the critical infrastructure components needed to enable AI agents to operate as independent economic participants in digital markets. The structure of the paper is as follows. Section 2 explores how code generation enables agents to interact programmatically with digital services—this capability transforms them from passive tools into potential economic participants. Section 3 analyzes how markets would benefit from AI agents as economic actors and presents a vision of digital economies where agents can innovate, create value, and participate in complex market interactions. These new market dynamics demand new infrastructure, which we examine in Section 4: authentication systems to establish agent identities, payment networks for machine-speed transactions, service discovery mechanisms for agent-readable capabilities, and standardized interfaces for agent interactions. The implications of this infrastructure extend beyond practical deployment, suggesting new paths to artificial intelligence through emergent market behaviors rather than monolithic systems, which we discuss in Section 5.

\section{The Bridge to Action}
Code generation has been one of the most successful applications of large language models, fundamentally changing how software is written\cite{li2022automatingcodereviewactivities}\cite{replit2024agent}\cite{cursor}\cite{windsurf}. While current tools like GitHub Copilot\cite{copilot} focus on assisting human developers, the true potential of code generation lies in enabling autonomous digital operation. Unlike traditional software that executes predefined instructions, AI systems can now write their own code to accomplish tasks, discover and integrate with new services, and adapt their capabilities based on results\cite{voyager}\cite{zhu2023ghost}. This shift from assisting humans to independent operation represents a fundamental change in how software interacts with digital systems.

The adoption of code generation tools has been remarkably rapid, with millions of developers now relying on AI assistants for their daily programming tasks\cite{github2024survey}. These systems excel not just at writing new code, but at understanding and modifying existing codebases - suggesting changes, implementing new features, and adapting code to match project patterns. Their success stems from large language models' ability to learn programming patterns, API usage, and common software design practices from vast amounts of public code. While they can struggle with complex codebases or sophisticated language features, requiring human guidance to navigate system-wide implications, early signs of autonomous behavior are emerging. Some systems now execute code, analyze test results, and iteratively refine their solutions based on errors and outputs, creating a primitive feedback loop\cite{replit2024agent}\cite{windsurf}. Yet these capabilities remain largely confined to specific tasks and bounded contexts - the high-level goals, constraints, and success criteria still come from human developers rather than emerging from the system's own objectives.

The shift toward true code agency emerges when AI systems generate and execute code to achieve their own objectives. Unlike current tools that respond to developer requests, autonomous systems can decompose high-level goals into concrete programming tasks, chain multiple steps together, and learn from execution results. An agent tasked with analyzing market data, for instance, might independently discover relevant APIs, write code to collect and process data, handle authentication and rate limits, and adapt its approach based on the results. When errors occur or APIs change, the system can diagnose issues, modify its code, and try alternative approaches - all without human intervention.

This autonomous code generation enables systems to expand their own capabilities by discovering and integrating new services. Rather than being limited to predefined functions, agents can read API documentation, understand service capabilities, and write code to incorporate them into their operations. This creates a form of digital embodiment where code generation serves as the bridge between understanding what needs to be done and actually doing it. The ability to write and execute code transforms these systems from passive responders to active participants in the digital world.

This capability to independently write, execute, and adapt code marks a fundamental transition in software systems. Where traditional programs are constrained by their initial design and human-written code, these new systems can dynamically extend their capabilities by integrating new services and creating novel combinations of existing ones. As AI systems become increasingly capable of autonomous code generation, they transform from tools that help create software into independent operators that can navigate and act within digital environments. This shift enables them to become active participants in digital systems, capable of discovering opportunities, creating new services, and engaging in complex digital interactions independently of human oversight.

\section{Markets as Coordination Systems}
Markets are decentralized systems that coordinate complex activities through the interactions of independent actors. These actors make local decisions based on their knowledge, responding to price signals that emerge from supply and demand dynamics \cite{hayek1945knowledge}. Through competition and cooperation, markets efficiently allocate resources and foster innovation without central planning \cite{smith1776wealth}.

In today's markets, the primary actors are humans - whether operating as individuals, corporations, or other organizational forms. These human actors leverage their local knowledge and expertise to identify opportunities, make decisions, and create value. They compete for resources while simultaneously cooperating through trade and contracts. Market feedback mechanisms help them adjust their strategies based on success or failure.

Markets coordinate activity through price signals that emerge from supply and demand dynamics. When entrepreneurs identify opportunities, they direct resources toward potential solutions - with profits signaling successful value creation and losses indicating the need to redirect resources. This feedback loop serves as a distributed computation system, where prices aggregate information about scarcity and value from countless participants \cite{hayek1945knowledge}. Successful innovations get amplified through investment and imitation, while resources flow away from failed approaches, driving continuous improvement in resource allocation \cite{schumpeter1942capitalism}.

The brilliance of markets lies in their emergent intelligence—how they coordinate vast and intricate activities in a way that is "smarter" than any individual participant or centralized system could achieve \cite{schelling1978micromotives}. This decentralized coordination enables humanity to undertake extraordinary feats. Consider the example of building rockets and sending them to space. No single entity orchestrates every aspect of the process. The materials for the rocket are mined in one part of the world, refined in another, and assembled in facilities that rely on countless other industries—from power generation to computer chip manufacturing. Meanwhile, the workers involved are sheltered, fed, and transported through an interconnected network of businesses that operate independently, all responding to market incentives. This staggering level of complexity is achieved without any central authority dictating every detail.

Similarly, cities emerge and function through the spontaneous order of decentralized decision-making \cite{schelling1978micromotives}. No central planner ensures that every household has sofas, televisions, or cars, yet these items are almost universally present. Homes are built by construction companies that procure materials from various suppliers, while furniture manufacturers design and produce items to meet diverse tastes. All of this is accomplished through countless actors responding to localized needs and opportunities, coordinated by market signals rather than direct orders. The result is an intricate, living system that evolves and adapts to the needs of its inhabitants without any single organization overseeing the process.

What makes this emergent order so remarkable is its efficiency and adaptability. Markets harness the knowledge and creativity of millions of individuals, aggregating their decisions into a system that can solve problems, allocate resources, and foster innovation on a scale no centralized entity could hope to match. This decentralized intelligence is not just impressive—it is foundational to how societies advance and thrive \cite{arthur2009nature}. It is proof of the profound power of coordination through markets, where the whole truly becomes far greater than the sum of its parts.

AI agents are now capable of joining markets as new types of participants. To understand why this is significant, it's important to first define what it means to be a market participant. A market participant is any entity that can process information, make decisions, and act based on economic principles like supply and demand. These participants engage in competition and cooperation, leveraging their local knowledge to identify opportunities and respond to price signals \cite{bostrom2014superintelligence}.
AI agents differ from human actors in two fundamental ways. First, there are quantitative improvements that simply enhance existing capabilities:
\begin{itemize}
    \item \textbf{Processing Speed:} While humans make decisions on timescales of minutes to days, AI agents can analyze situations and respond in milliseconds \cite{lopez2018financial}.
    \item \textbf{Information Processing:} Agents can simultaneously process and analyze vastly more data points than humans \cite{kaplan2020scaling}.
    \item \textbf{Operational Continuity:} Unlike humans who require sleep and breaks, agents can operate continuously \cite{brynjolfsson2014second}.
    \item \textbf{Multi-tasking Capacity:} Agents can simultaneously engage in numerous market interactions \cite{brynjolfsson2014second}.
\end{itemize}

However, AI agents also possess unique features that set them completely apart from humans:
\begin{itemize}
    \item \textbf{Perfect Replication:} An agent's code, knowledge, and strategies can be perfectly copied across other instances without loss of information or effectiveness \cite{nisan2007algorithmic}.
    \item \textbf{Dynamic Instantiation:} Agents can be created or terminated instantly in response to market opportunities, enabling perfect market elasticity \cite{nisan2007algorithmic}.
    \item \textbf{Collective Learning:} Agents can share their knowledge and experiences perfectly through software, eliminating information silos \cite{malone2022handbook}.
    \item \textbf{Consistent Performance:} Agents maintain the same level of performance indefinitely, without variation due to fatigue or emotional factors \cite{malone2022handbook}.
    \item \textbf{Resource Fluidity:} While humans are limited to their individual brains, computational resources—like GPUs, for example—can be dynamically reallocated between agents \cite{marinescu2022cloud}.
\end{itemize}

These revolutionary differences introduce entirely new dynamics for market organization and operation that weren't possible in purely human markets.

These differences mean that AI agents don't simply replicate human market behavior, they augment and expand it. While traditional economic frameworks remain relevant for studying AI agents in markets, we must also develop new models to understand their unique dynamics and implications.

The potential of AI agents as economic actors is already evident, even before considering physical embodiment through robotics. By participating in digital markets, these agents can create significant value through automated trading, service optimization, resource allocation, and complex coordination tasks \cite{lopez2018financial}. Their ability to operate continuously, process vast amounts of information, and perfectly share successful strategies enables new forms of economic organization and value creation that were previously impossible.

However, these theoretical capabilities face practical constraints in today's digital infrastructure. Current systems are built around human interaction patterns—they assume actors need rest, operate at human speed, and make decisions through human interfaces. To unlock the full potential of AI agents in markets, we must address these infrastructure limitations and create systems that can support their unique capabilities and dynamics \cite{brynjolfsson2011race, greenstein2019digital}.

\section{Infrastructure Challenges}

Imagine an entrepreneur, Sam, who while working at an online retailer notices that their reporting tools aren't effectively tracking how promotional campaigns impact customer lifetime value. Working with the sales team, he builds a simple dashboard that combines marketing spend and customer purchase history in a way that clearly shows which promotions lead to long-term customer relationships. The sales team loves it, and several other companies express interest when he demonstrates it at an industry conference. Realizing there's clear market demand, he decides to turn this solution into a Software-as-a-Service (SaaS) product.

What might seem like a straightforward path — packaging his dashboard into a service — quickly reveals the complex web of digital infrastructure and third-party services essential to building a modern software business. To turn his solution into a valuable product, Sam needs robust cloud infrastructure to process data reliably, a secure system for managing customer accounts and data, and a seamless way to handle subscriptions and payments.

Sam's journey begins with registering a domain name. He visits a domain registrar's website where he browses available domains by typing potential names into a search bar. After finding an available domain, he clicks through a checkout process that requires creating an account. He provides his email, chooses a password, and enters his contact details. To complete the purchase, he inputs his credit card information and verifies his identity through a code sent to his phone.

Next, he needs to set up his cloud infrastructure. After comparing providers through their websites and documentation, he signs up for an account, again providing business details and payment information. He navigates through web consoles to provision servers, set up databases, and configure networking rules \cite{marinescu2022cloud}. Each step involves clicking through interfaces designed for human operators, reading documentation, and making configuration choices through forms and dropdown menus.

For his landing page, Sam uses a website builder that lets him design through a visual interface. He connects analytics tools by copying and pasting tracking codes, following step-by-step tutorials in their documentation. Each service required Sam to create an account, verify his identity, and maintain state through human-oriented dashboards \cite{cameron2005laws}. Even when he needed programmatic access to services, he first had to work through human touchpoints—creating accounts through web forms, verifying his identity, and studying documentation. Payment processing similarly revolves around credit card workflows designed for human cardholders, with verification steps that assume and require human input \cite{haldane2008future}.

Each of these steps brings Sam closer to launching his Minimum Viable Product (MVP). What started as a clever insight into promotional campaign effectiveness now requires an ecosystem of supporting services to reach its customers: authentication to secure customer data, analytics to understand feature usage, monitoring tools to ensure reliability, email systems to communicate with users, and payment infrastructure to capture revenue. While Sam's dashboard remains the core innovation, delivering it as a modern SaaS product means wrapping it in layers of digital infrastructure that other companies have built and maintained \cite{brynjolfsson2011race}. This pattern of building on existing services reveals fundamental assumptions about how software is discovered, accessed, and integrated in today's digital economy.

Sam's story illustrates how deeply human-centric assumptions are embedded in our current digital infrastructure. Every layer of interaction—from discovering services to consuming them and processing payments—is built around human operators making decisions and taking actions \cite{dix2004human}.

Service discovery relies on methods natural to humans: insights gained at industry conferences, browsing websites, and following recommendations. Each service presents itself through carefully crafted landing pages, documentation, and pricing tables meant to be evaluated by human eyes and reasoning. Even modern discovery mechanisms like search engines and app stores are optimized for human browsing patterns.

The consumption of these services follows an even more rigid human-centric pattern. Every interaction assumes a human operator navigating through browser-based interfaces or mobile applications. Each service required Sam to create an account, verify his identity, and maintain state through human-oriented dashboards. Even when he needed programmatic access to services, he first had to work through human touchpoints—creating accounts through web forms, verifying his identity, and studying documentation. Payment processing similarly revolves around credit card workflows designed for human cardholders, with verification steps that assume and require human input.

Each of these processes contains fundamental blockers for machines that must be solved to enable AI agents to participate as normal economic actors. Service discovery today relies heavily on real-world networks—conference presentations, colleague recommendations, and advertisements designed for human attention. Account creation and payment processes actively resist automation through CAPTCHA systems\cite{singh2014survey} and other anti-bot measures \cite{douceur2002sybil}. These measures were historically necessary to defend against malicious automation but now block legitimate AI participants. In the following sections, we'll examine each of these challenges in detail and explore promising solutions that could enable AI agents to become full participants in the digital economy.

\section{Service Discovery}
\label{sec:service_discovery}

\input{figures/service_discovery_mindmap}

Success in markets depends critically on information - not just about prices and competition, but about what solutions already exist. No business operates in isolation. From manufacturing to software development, the most successful enterprises build upon existing solutions rather than reinventing every component from scratch. This specialization and reuse of existing capabilities is fundamental to market efficiency and innovation.

Consider a modern software company building a new service. Rather than implementing their own payment processing, email delivery, authentication system, and cloud infrastructure, they typically assemble these capabilities from existing providers. This allows them to focus their resources and creativity on their core innovation - the unique value they bring to market. The same pattern repeats across industries: automotive manufacturers source components from specialized suppliers, restaurants rely on established food distribution networks, and retailers build on existing logistics infrastructure.

This reliance on existing solutions creates a critical challenge: how do market participants discover what's available? Before any economic decision can be made - whether selecting a supplier, choosing a service provider, or identifying market opportunities - participants must first become aware of their options. The efficiency of markets thus depends not just on price signals, but on the mechanisms through which participants discover and evaluate potential solutions. In human markets, these discovery mechanisms have evolved over centuries, from trade fairs to modern digital platforms. As AI agents begin to participate in markets, we must examine whether these mechanisms will serve their needs or whether new approaches are required.

\subsection{Current Infrastructure}
\label{subsec:discovery_current}
Information about available market solutions reaches participants through multiple complementary channels. The most fundamental is experiential discovery - learning through daily life and professional activities. Market participants naturally become aware of solutions through their work, observing what others use, attending industry events, and participating in professional communities. This ambient awareness forms a foundation of market knowledge that shapes future decisions and investigations.

Social discovery builds upon this experiential base through professional networks. When faced with a need, participants often turn first to trusted colleagues and peers, seeking recommendations based on direct experience. These recommendations carry particular weight because they come with context and validation from trusted sources. Communities, both formal and informal, serve as repositories of collective knowledge about what solutions exist and how well they work in practice.

Beyond passive channels, participants actively seek out information through traditional marketing and industry channels. Trade publications, conferences, and advertising campaigns help market participants stay informed about available solutions and innovations. These mechanisms not only announce the existence of solutions; they help participants understand capabilities, compare options, and evaluate fit for their needs.

The digital revolution introduced new discovery mechanisms that operate at unprecedented scale. Search engines index vast amounts of information about services and solutions, while specialized directories and review platforms aggregate structured data about specific market segments. These digital channels differ fundamentally from their predecessors - they are programmatically accessible, operate continuously, and can process queries at machine speed. Yet they were still designed primarily for human consumption, with interfaces and information architecture optimized for human cognitive patterns.

\subsection{Limitations for AI Agents}
\label{subsec:discovery_limitations}
The discovery methods that serve human participants so well face fundamental limitations when applied to AI agents. Most critically, agents currently lack physical embodiment that enables experiential discovery. They cannot attend conferences, engage in water cooler conversations, or observe solutions in use during their daily activities. This absence of real-world presence cuts them off from the rich stream of ambient information that humans unconsciously process and integrate into their market awareness.

Even digital discovery channels, despite being technically accessible to agents, present significant challenges. Marketing materials and service documentation are optimized for human consumption, leveraging psychological patterns and visual presentation that may not translate meaningfully to machine understanding. Landing pages use persuasive design, emotional appeals, and carefully crafted imagery - techniques honed over decades to influence human decision-making but largely irrelevant to machine evaluation of capabilities.

Information about services is typically fragmented across multiple sources in ways that reflect human discovery patterns. While agents can process vast amounts of information quickly, current presentation formats may cause them to miss relevant capabilities or connections. A service's value proposition might be spread across marketing materials, technical documentation, and user testimonials - an organization that follows human information consumption patterns but may not be optimal for machine discovery and evaluation.

While agents can technically access existing discovery channels, doing so is often inefficient and cumbersome. They must parse human-oriented interfaces, extract relevant information from unstructured content, and navigate multiple systems designed around human workflow assumptions. This creates unnecessary friction in the discovery process, reducing the potential speed and efficiency gains that agent-based market participation could offer.

\subsection{Future Design Considerations}
\label{subsec:discovery_future}
The limitations of current discovery mechanisms point toward new infrastructure designs that could better serve AI agents as market participants. Service registries could provide machine-readable descriptions of capabilities, pricing, and integration requirements in standardized formats. These descriptions would embed the complete context needed for evaluation in a single request, allowing agents to assess potential solutions without navigating multiple information layers.

Several approaches could emerge to facilitate discovery between services and agents. Existing indexing infrastructure like search engines and service directories could evolve to better serve agent discovery needs. Rather than focusing on keyword matching and human-readable content ranking, these systems could develop specialized capabilities for understanding and exposing service characteristics, integration requirements, and operational constraints. This might involve new crawling strategies optimized for machine-readable service descriptions, index structures that facilitate capability-based matching, and query interfaces designed around agent decision patterns. Alternative approaches draw inspiration from distributed systems, where gossip protocols efficiently propagate information across networks.

Machine-friendly discovery mechanisms could also leverage agents' unique information processing capabilities. Instead of progressive disclosure models designed for human attention spans, these systems could provide comprehensive technical specifications and integration requirements upfront. Other possibilities include semantic service descriptions, capability-based discovery protocols, or real-time service meshes that dynamically match agent requirements with available solutions. The key lies in designing systems that align with how agents process and evaluate information while maintaining compatibility with existing infrastructure.

Another thing to keep in mind is that the emergence of AI-powered services introduces unprecedented flexibility in how capabilities are discovered and matched. Unlike traditional services with fixed feature sets, AI-enabled services may adapt their capabilities dynamically based on consumer needs. This flexibility extends to how services present themselves - rather than maintaining static descriptions, services could engage in dynamic capability negotiation, expressing their potential value in context-specific ways. Furthermore, the speed of AI development suggests markets where services evolve, merge, or become obsolete at machine speed rather than human timescales. Discovery mechanisms must therefore handle not just static service descriptions but continuous streams of capability updates, service transitions, and market reconfigurations.

\section{Identity and Authorization}
Digital services fundamentally rely on knowing who or what is making requests and what they're allowed to do. Every API call, every database query, and every transaction must address two critical questions: "Who are you?" (authentication) and "What are you allowed to do?" (authorization). While these might seem like simple security concerns, they form the foundation that enables stateful services to operate, track usage, manage resources, and maintain consistency across interactions. Authentication establishes identity—proving that an entity is who it claims to be—while authorization determines what authenticated entities can do, controlling their permissions and access rights within the system.

The landscape of digital identity has evolved far beyond simple human-to-service interactions. Modern systems must handle complex scenarios where software acts on behalf of other software, requiring sophisticated delegation mechanisms and permission models. These interactions must maintain security while providing audit trails for compliance, tracking resource usage for billing, and establishing verifiable chains of trust between different services and systems.

In this section, we examine the infrastructure that enables identity and authorization in digital services. We begin by analyzing current methods and understanding why they evolved to their present form. We then explore how these systems, built around human-centric assumptions, present fundamental limitations for AI agents operating at machine speed and scale. Finally, we investigate promising approaches that could better serve the needs of autonomous agents while maintaining security and accountability. These considerations are critical as we move toward digital systems where AI agents become primary participants rather than occasional automated actors.

\subsection{Current Infrastructure: Identity}
\label{subsec:identity_current}

\input{figures/identity_mindmap}
    
Today's digital identity infrastructure evolved primarily around human users and their needs, creating a hierarchy of solutions from simple authentication to complex federated systems. At its most basic level, username and password combinations establish digital identity through shared secrets. While this approach remains widespread, it presents significant operational challenges - password reuse, weak choices, and credential theft create security risks that complexity requirements and rotation policies only partially mitigate.

For larger organizations, federation protocols extend this basic model across organizational boundaries while maintaining local control. Standards like SAML enable centralized identity management with distributed verification - an employee can use their corporate credentials across multiple internal systems without creating separate accounts. This creates a natural hierarchy where identities can be established at one level (like an organization) and inherited or delegated to lower levels (like departments or individual services).

As the internet grew and services needed to interact with previously unknown parties, Public Key Infrastructure (PKI) and certificate-based systems emerged as a solution for establishing cryptographic proof of identity. Unlike username/password or federated systems which operate in closed environments, PKI enables secure communication between parties that have never interacted before. Certificate authorities act as trusted intermediaries, validating identities and issuing certificates that create verifiable chains of trust. The Automated Certificate Management Environment (ACME) protocol made PKI practical at scale, underpinning critical security infrastructure like HTTPS.

For service-to-service communication within controlled environments, API keys provide a more streamlined solution. These long-lived tokens enable programmatic authentication with clear scoping and audit capabilities. Unlike passwords, API keys can maintain sufficient entropy to resist attacks while supporting automated rotation and revocation. This makes them particularly suitable for automated systems that need to maintain clear records of which keys are used for which operations.

Authentication mechanisms have also evolved to account for the physical world and human factors. Two-factor authentication adds a second verification layer through physical devices like YubiKeys or phone-based authenticators. These systems recognize that digital identity must often bridge the gap between the online and offline worlds, especially for high-value operations or sensitive data access.

Session management adds another layer to this ecosystem. While identity establishes who someone is, authentication mechanisms like JSON Web Tokens (JWTs) allow services to efficiently verify that a user has already proven their identity without repeating the full authentication process. Modern API gateways use these tokens to maintain authenticated sessions, validating requests without having to re-verify credentials for each operation.

All these systems share common patterns for establishing, verifying, and revoking identities. Identity establishment typically happens infrequently, with careful verification steps and approval processes. Identity verification occurs much more frequently but relies on caching and pre-established trust relationships. Revocation follows similar patterns, with certificate expiration and rotation measured in days or months rather than seconds or milliseconds.

The audit requirements around identity also reflect these operational patterns. Systems maintain logs of identity creation, modification, and usage that assume relatively stable identities performing discrete actions. This allows for after-the-fact investigation and compliance verification, with audit trails typically focusing on exceptional events rather than routine operations.

This infrastructure has proven remarkably successful for its intended use cases, but its assumptions about operational pace and scale remain fundamentally aligned with human timescales and physical world constraints. From manual access approval to session lifetimes and identity change rates, these deeply embedded patterns form the foundation of today's digital trust and authentication mechanisms.

\subsection{Limitations for AI Agents: Identity}
\label{subsec:identity_limitations}
The identity infrastructure we've described, while robust for human users and traditional services, faces fundamental challenges when applied to AI agents. These limitations stem not just from issues of scale, but from core assumptions about how digital identities behave and interact.

Scale presents the most immediate challenge. While current systems can handle millions of human users, they assume these identities remain relatively stable over time. AI agents, in contrast, might be created and destroyed thousands of times per second as they adapt to changing conditions. Certificate authorities and identity verification systems designed for human-scale operations become bottlenecks when agents need new credentials at machine speed. Even automated systems like ACME protocols, which can issue certificates in minutes, operate far too slowly for agents that might exist for milliseconds.

The relationship between identity and physical reality also breaks down. Traditional systems often rely on bridging digital and physical worlds - two-factor authentication assumes access to physical devices, identity verification often requires government documents or biometric data, and high-stakes operations might demand in-person verification. AI agents exist purely in the digital realm, making these physical anchors simultaneously irrelevant and problematic since many security systems require them.

Identity hierarchies face similar challenges. In human organizations, identity hierarchies mirror organizational structures - employees inherit permissions from their departments, contractors receive limited delegated access, and these relationships change infrequently. AI agents, however, might spawn hierarchies of sub-agents dynamically, each needing its own identity while maintaining cryptographic proof of its relationship to its parent. Traditional federation protocols weren't designed for these rapidly evolving, deeply nested identity relationships.

Trust establishment becomes particularly challenging with ephemeral agents. Traditional systems rely on long-lived identities building reputation over time - a user's account history, an organization's business registration, or a service's track record all contribute to trust. But how do you establish trust for an agent that might exist for milliseconds? How do you maintain reputation systems when identities are fluid and short-lived? Current infrastructure provides no clear answers for establishing and verifying trust between autonomous systems operating at machine speed.

Audit requirements compound these challenges. Traditional audit trails assume relatively stable identities performing discrete actions that can be logged and reviewed. When agents create and destroy thousands of child identities per second, each performing hundreds of operations, traditional logging approaches become impractical. Yet the need for accountability and forensic analysis remains, especially when agents operate with real-world consequences.
Current identity infrastructure, optimized for human-scale operations and physical world anchors, becomes a fundamental limiting factor in enabling truly autonomous agent interactions.

\subsection{Future Design Considerations: Identity}
\label{subsec:identity_future}
The limitations of current identity systems for AI agents demand new approaches that maintain security and accountability while operating at machine speed. Several promising directions emerge when we rethink fundamental assumptions about digital identity.

Public key cryptography offers a foundation for addressing the scale challenge. By allowing agents to generate their own key pairs and prove ownership cryptographically, systems can enable autonomous identity creation at machine speed without centralized bottlenecks. Rather than waiting for certificate authorities or federation servers, agents can create verifiable identities instantly while maintaining cryptographic proof of their authenticity.

The traditional account creation and login flow also needs reimagining for AI agents. While standards like WebAuthn and passkeys are eliminating passwords for human users, new protocols could emerge specifically for machine-to-machine account creation. These could allow agents to programmatically establish service relationships through cryptographic attestation rather than traditional registration flows. Instead of filling out forms and verifying emails, agents could prove their capabilities and trustworthiness through cryptographic challenges, enabling instant service onboarding while maintaining security.

For ephemeral identities and rapid trust establishment, zero-knowledge systems suggest promising approaches. These systems enable agents to prove properties about themselves - their authorization level, their operational history, their delegation chain - without revealing unnecessary details. Combined with verifiable credentials, an agent could prove it was spawned by a trusted parent or has performed similar operations successfully, enabling a meaningful reputation even for short-lived identities.

The challenge of identity hierarchies and dynamic delegation could be addressed through capability-based systems. These allow an agent to delegate subsets of its identity and permissions to child agents while maintaining cryptographic proof of the delegation chain. When combined with attribute-based systems, this enables dynamic, context-aware identity verification based on provable properties rather than static relationships.
To handle the separation from physical anchors, new trust models could emerge based on observed behavior and computational proof rather than real-world verification. Decentralized reputation systems could track agent behavior across short lifespans, while proof-of-computation systems could verify an agent's capabilities and intentions through demonstrated work rather than external validation.

The path forward likely involves combining multiple approaches, creating layered systems that can handle both long-lived and ephemeral identities while maintaining security and accountability at machine speed. As AI agents become more prevalent in digital systems, solving these identity challenges becomes crucial for enabling their full participation in digital markets.

\subsection{Current Infrastructure: Authorization}
\label{subsec:authorization_current}

\input{figures/authorization_mindmap}

Once a system knows who is making a request, it must determine what that entity is allowed to do. Authorization systems have evolved to manage these permissions at scale, balancing security with operational efficiency.

Role-Based Access Control (RBAC) has traditionally been the dominant model for managing permissions in large systems. Rather than assigning permissions directly to users, RBAC groups related rights into roles that map to organizational functions or job titles. This abstraction simplifies permission management - instead of tracking individual permissions across thousands of users, administrators can assign roles like "admin," "editor," or "viewer."

More recently, Relationship-Based Access Control (ReBAC) has emerged as a powerful complement to RBAC, particularly for applications with complex social or organizational relationships. ReBAC determines permissions based on how entities relate to each other within the system. For example, in a document management system, users might be able to edit documents owned by their direct reports or view documents from anyone in their department. This model naturally captures real-world permission patterns that are cumbersome to express in traditional RBAC.

These authorization decisions are typically enforced through tokens. Stateful approaches use session identifiers stored server-side, often in cookies, allowing for immediate permission revocation but requiring central state management. Stateless approaches using JSON Web Tokens (JWTs) encode the authorization information directly in the token, enabling faster verification but making revocation more challenging. Systems often combine both approaches, using short-lived JWTs with periodic refreshes from a stateful system.

For cross-service scenarios, OAuth has become the standard protocol for delegating access rights across organizational boundaries. OAuth enables controlled access sharing without credential exchange - a user can allow one service to access their data on another service without sharing their password. Different OAuth flows serve different use cases: the authorization code flow handles web applications, while the client credentials flow enables service-to-service communication. The protocol's separation of authentication and authorization concerns has made it particularly suitable for modern distributed systems.

Most current authorization systems operate with relatively coarse-grained permissions that change infrequently. A typical enterprise might update role definitions monthly or quarterly, with individual permission changes happening daily or weekly. This relatively slow pace allows for manual review of permission changes and simplifies auditing and compliance tracking.

\subsection{Limitations for AI Agents: Authorization}
\label{subsec:authorization_limitations}
Current authorization systems, designed around human organizational structures and workflows, face significant limitations when applied to AI agents operating at machine speed and scale.

The core challenge stems from how permissions need to be evaluated. Traditional RBAC systems work well when roles change infrequently and map cleanly to organizational hierarchies. However, AI agents might need to adjust their permissions thousands of times per second based on their current task or context. While ReBAC better captures relationship-based permissions, current implementations aren't designed to handle relationships that form and dissolve at machine speed.

Cross-service authorization presents particular challenges. OAuth works well for relatively stable delegation patterns, but AI agents might need to establish and revoke delegated access continuously as they spawn child agents or collaborate on tasks. The overhead of traditional OAuth flows becomes prohibitive when operating at machine timescales. Additionally, the standard OAuth scopes are too coarse-grained for agents that need precise, task-specific permissions.

Context-aware access control becomes crucial yet problematic for AI agents. An agent's permissions might need to change based on its current task, the data it's processing, system load, or other environmental factors. Traditional authorization systems aren't designed to incorporate this rich context into real-time permission decisions. While some systems support basic contextual rules, they typically can't handle the complex, dynamic conditions that govern AI agent behavior.

Token-based authorization systems face their own challenges with AI agents. Stateful tokens require central storage that becomes a bottleneck at machine scale. Stateless tokens like JWTs, while more scalable, make it difficult to revoke permissions quickly when agent behavior or system conditions change. The traditional compromise of short-lived tokens with refresh mechanisms introduces latency that impacts agent operations.

These limitations compound each other in practice. An AI agent might need to spawn multiple child agents, each requiring specific permissions based on their task and context while coordinating access across multiple services - all at machine speed. Current authorization systems, optimized for human-scale operations with relatively static permissions, become a fundamental bottleneck in enabling truly autonomous agent interactions.

\subsection{Future Design Considerations: Authorization}
\label{subsec:authorization_future}
The limitations of current authorization systems for AI agents require new approaches that can handle dynamic, context-aware permissions at machine speed while maintaining security. Several promising directions emerge when we rethink how permissions should work in agent-driven systems.

Capability-based security models offer a foundation for handling fine-grained permissions at scale. Systems like macaroons and biscuits enable unforgeable tokens that can encode complex permissions and delegation rights. These tokens can be attenuated - an agent can create restricted versions of its own capabilities for child agents - while maintaining cryptographic proof of the delegation chain. This enables secure permission delegation without requiring central coordination.

Context-aware authorization could be achieved through real-time policy evaluation engines optimized for machine-speed decisions. Rather than static role assignments, these systems would evaluate permissions based on current conditions, agent behavior, and system state. By encoding authorization logic in verifiable, deterministic rules, agents could even predict whether they would have necessary permissions before attempting operations.

Cross-service authorization could evolve beyond traditional OAuth flows through standardized permission protocols designed for agent-to-agent interactions. These would enable rapid establishment and verification of permissions across service boundaries while maintaining security. Services could publish their permission models in machine-readable formats, allowing agents to automatically discover and request necessary access rights.

The infrastructure supporting these approaches needs to handle massive scale without sacrificing security. This suggests architectural patterns where permission verification happens as close to the edge as possible, with cryptographically verifiable tokens that can be evaluated without central coordination. Audit trails would need to efficiently track permission changes and usage without becoming a bottleneck.

While these approaches show promise, they require careful design to prevent abuse. Capability-based systems must prevent unauthorized escalation of privileges. Context-aware systems need protection against gaming or manipulation of the context. Cross-service protocols must maintain security even when some services are compromised. As AI agents become more prevalent in digital systems, solving these authorization challenges becomes crucial for enabling their safe participation in digital markets.

\section{Software Interfaces}
\label{sec:software_interface}

\input{figures/software_interfaces_mindmap}

Software service consumption has undergone a dramatic evolution over the past decades, transforming from simple command-line interfaces into rich, multi-modal experiences. This evolution reflects both advancing technical capabilities and our deepening understanding of human-computer interaction patterns.

The earliest software interfaces were purely text-based, requiring users to memorize specific commands and syntax. As graphical user interfaces emerged, they introduced new paradigms built around visual metaphors - windows, icons, menus, and pointers. These interfaces made software more accessible by mapping complex operations to intuitive visual elements that users could manipulate directly.

The web browser represents perhaps the most significant shift in how we consume software services. What began as a simple document viewer has evolved into a universal application platform. The browser's combination of standardized technologies, built-in security model, and instant access to services without installation has made it the dominant platform for software delivery. Today, many applications that were once exclusively desktop software have migrated to browser-based versions, from productivity tools to complex enterprise systems.

While the browser dominates, software consumption isn't monolithic. Desktop applications remain important for compute-intensive tasks or deep operating system integration. Mobile applications offer optimized experiences for smaller screens and touch interfaces. Many modern services support multiple consumption patterns - they might offer a web interface for human users, native mobile apps for on-the-go access, and APIs for programmatic integration.

This diversity of interfaces reflects a fundamental truth: how we consume software shapes what's possible with it. The interface is not just a wrapper around functionality - it defines the patterns of interaction, sets expectations about response times and data presentation, and ultimately constrains how value can be extracted from the service. As we move toward a future where AI agents become active consumers of software services, these patterns and constraints take on new significance.

\subsection{Current Infrastructure}
\label{subsec:interfaces_current}
Today's software services are primarily consumed through two distinct patterns: user interfaces designed for human interaction, and programmatic interfaces designed for machine-to-machine communication. Each pattern has evolved its own conventions, constraints, and optimization strategies.

\subsubsection{User Interface (UI)}
User interfaces dominate human interaction with software services. Whether through web browsers, desktop applications, or mobile apps, these interfaces share common characteristics shaped by human cognitive and perceptual abilities. They present information visually, often breaking complex data into manageable chunks spread across multiple screens or views. They rely on progressive disclosure - showing basic information first with the option to drill deeper - to avoid overwhelming users.

The web browser has emerged as the primary platform for delivering user interfaces. Its ubiquity, built-in security model, and ability to update instantly make it ideal for modern service delivery. Web applications use HTML, CSS, and JavaScript to create rich interactive experiences that work across devices. The browser's standardized technologies and APIs provide a consistent foundation for building complex applications that once required native desktop installation.

Modern UI design patterns reflect a deep understanding of human information processing limits. Navigation structures, form layouts, and data visualization techniques are all optimized around human perceptual capabilities and attention spans. Even seemingly simple choices like the number of items displayed in a list or the depth of a menu structure are calibrated to human cognitive load limits.

\subsubsection{Application Programming Interface (API)}
Alongside user interfaces, most modern services offer application programming interfaces (APIs) for machine-to-machine communication. These interfaces expose service functionality in ways that other software can consume directly, without human intervention. REST and GraphQL have emerged as dominant paradigms for API design, offering structured ways to request and manipulate data.

APIs typically exchange data in formats like JSON or XML that balance human readability with machine parsing. They implement authentication mechanisms, rate limiting, and usage quotas to manage resource consumption. Many services provide software development kits (SDKs) that wrap their APIs in language-specific libraries, making it easier for developers to integrate services into their applications.
11
While programmatic interfaces enable automation, they're still largely designed around human development patterns. API designs prioritize clarity and ease of understanding over machine efficiency. The documentation assumes human readers who can interpret examples and understand the context. Even rate limits and quotas are typically set based on expected human-driven usage patterns rather than machine capabilities.

This dual infrastructure - visual interfaces for humans and programmatic interfaces for machines - has served well for traditional software integration needs. However, as AI agents emerge as a new class of software consumers, the assumptions built into both patterns face new challenges.

\subsection{Limitations for AI Agents}
\label{subsec:interfaces_limitations}
Current service consumption patterns, optimized for either human interaction or traditional machine-to-machine communication, present several fundamental challenges for AI agents.

UI-based consumption poses immediate challenges for AI agents. Modern interfaces are built around human visual processing capabilities and cognitive patterns. Elements like buttons, forms, and navigation menus rely on visual recognition and spatial relationships that make perfect sense to humans but require complex interpretation by machines. While browser automation tools can interact with these elements, they must essentially simulate human interaction patterns rather than engaging with the underlying functionality directly. The human-centric design of web interfaces also creates inefficiencies in data access. Information that could be transmitted in a single response is often spread across multiple pages or views to avoid overwhelming human users. What a human experiences as a natural flow - clicking through pages of search results or navigating through hierarchical menus - becomes a series of forced sequential operations for an AI agent capable of processing far more information in parallel.

While APIs might seem better suited for AI agent consumption, current implementations present their own challenges. REST and GraphQL interfaces, while structured, still rely heavily on human-readable formats and human-oriented usage patterns. The overhead of parsing JSON or XML, while negligible for occasional requests, becomes significant when scaled to thousands of operations per second.
Rate limiting and throttling mechanisms, designed around human-scale usage patterns, can severely constrain AI agents' ability to operate efficiently. These limits often assume traditional software integration patterns rather than the high-frequency, parallel operations that AI agents might need to perform.

Perhaps most fundamentally, current service consumption patterns enforce artificial constraints on how functionality can be accessed. An AI agent must either navigate human-centric UIs through automation or work within the boundaries of pre-defined APIs. There's no middle ground that would allow agents to dynamically discover and integrate with service capabilities at runtime.
The separation between UI and API consumption also creates inefficiencies. While a human user might benefit from rich visual feedback and progressive disclosure, an AI agent could process the same information more efficiently in a single structured response. Yet services rarely offer this kind of flexible consumption model that could adapt to the consumer's processing capabilities.
These limitations mean that AI agents often must operate far below their theoretical capabilities, constrained by infrastructure designed for different patterns of consumption. As agents become more sophisticated, these constraints will increasingly become bottlenecks to their effective operation in digital systems.

\subsection{Future Design Considerations}
\label{subsec:interfaces_future}
Creating infrastructure that better supports AI agent consumption of services requires rethinking fundamental assumptions about how software functionality is exposed and accessed. Several promising directions emerge when we consider the unique capabilities and requirements of AI agents.

At the protocol level, new standards could emerge specifically for agent-service interaction, optimized for machine-speed operations and parallel processing. These protocols might support dynamic capability negotiation and efficient data formats, moving beyond current REST and GraphQL paradigms. Enhanced RPC frameworks could support more flexible calling patterns while maintaining the performance benefits of traditional RPC, enabling efficient machine-to-machine communication that better matches how AI agents operate.

Services could expose their capabilities through rich, machine-readable descriptions that go beyond traditional API documentation. These descriptions would enable AI agents to understand not just available endpoints, but the semantic meaning of operations, their prerequisites, and their effects. Service providers could implement AI interfaces that act as intelligent intermediaries, understanding natural language queries and translating them into appropriate internal operations. This approach would allow services to maintain their existing infrastructure while providing a more intuitive interface for AI agents.

Rather than maintaining separate UI and API interfaces, services might adopt unified interfaces that adapt to the consumer's capabilities. These interfaces could adjust their response format and granularity based on whether they're dealing with a human user or an AI agent, eliminating the current inefficiencies of forcing agents to either navigate human UIs or work within constrained API boundaries. These adaptive interfaces would support both traditional consumption patterns and new agent-oriented interactions.

The infrastructure could also support dynamic service composition, allowing agents to discover and combine service capabilities at runtime. This would enable agents to create new workflows and applications by combining existing services in novel ways, operating at machine speed and scale. By supporting this kind of dynamic integration, the infrastructure would enable entirely new patterns of service consumption and value creation.

These enhancements would fundamentally change how AI agents interact with software services, enabling them to operate more efficiently and create more value through dynamic service integration. However, implementing these changes requires careful consideration of security, stability, and backward compatibility with existing systems. The transition to this new infrastructure will likely be gradual, with services initially offering enhanced capabilities alongside traditional consumption patterns.

\section{Payments}
\label{sec:payments}

\input{figures/payments_mindmap}

Payment systems were notably absent from the internet's original design. As e-commerce emerged, existing payment networks like Visa and Mastercard were adapted to work online. This adaptation required significant infrastructure changes to secure payment data transmission. The payment card industry developed extensive security standards (PCI DSS) to protect sensitive financial information, requiring merchants to maintain secure systems and undergo regular audits. These requirements created substantial overhead for accepting online payments.

Companies like Stripe later simplified this complexity by providing APIs that handle payment processing and PCI compliance, making it easier for digital services to accept card payments. Mobile payment solutions like Apple Pay and Google Pay built additional layers on top of these card networks, improving user experience while still ultimately relying on the same underlying infrastructure.

Digital payment platforms like PayPal introduced an alternative approach by creating closed ecosystems where users maintain account balances. These systems enable faster transfers by simply updating internal ledgers rather than initiating traditional bank transfers for each transaction. While efficient within their networks, these platforms essentially operate as isolated silos.

The need for internet-native payments was recognized early - the HTTP specification even included a status code 402 "Payment Required." This acknowledgment that the web would need native payment capabilities proved prescient, though the code remains largely unused as no standard implementation emerged.

Earlier attempts to create digital cash predated e-commerce itself. In the 1980s, David Chaum developed eCash, a system designed to provide anonymous electronic transactions. While technically sophisticated, eCash and similar systems required backing from traditional financial institutions and failed to achieve widespread adoption.

The introduction of Bitcoin in 2008 presented a different approach to electronic money - one that operated independently of existing financial institutions. This innovation has grown into a significant financial infrastructure, particularly for cross-border value transfer. The ecosystem has expanded to include stablecoins like USDC and USDT, which maintain price stability by being backed with traditional currency reserves, bridging the gap between traditional finance and digital assets.

In this section, we examine how the current payment infrastructure serves digital markets, its limitations for AI agents, and potential architectures that could enable autonomous economic participation.

\subsection{Current Infrastructure}
\label{subsec:payments_current}
Traditional payment networks form the backbone of online commerce. When a customer makes a credit card purchase, their card details are captured through secure browser forms and immediately tokenized - converting sensitive data into secure tokens that can be stored and reused without exposing the actual card numbers. The transaction involves multiple parties: the issuing bank (customer's bank), the acquiring bank (merchant's bank), and the card network (Visa/Mastercard) that connects them. The payment flow differs significantly between one-time payments and subscriptions. While one-time payments require explicit customer action, subscription systems operate on a "pull" model where merchants can automatically charge previously stored payment methods. This process involves two distinct phases: real-time authorization where the payment is approved, and settlement which typically occurs days later when funds actually move between banks. For international transactions, this system grows more complex, requiring correspondent banking relationships to move money across borders. Traditional bank transfers like ACH and wires still handle significant transaction volume, especially for business payments, though they operate on even slower settlement schedules.

Payment systems also integrate deeply with business operations beyond just moving money. Merchants must collect billing information for tax purposes, often connecting payment processing with accounting software for automated reconciliation and reporting. While these networks excel at handling larger transactions, they were not designed for micropayments. Instead, services typically implement metering or credit-based systems where users pre-purchase credits or are billed periodically based on accumulated usage.

Modern payment processors have innovated by abstracting this complexity. Rather than requiring merchants to establish their own merchant accounts and banking relationships, processors handle the entire payment stack. They provide unified APIs that support multiple payment methods while managing the underlying complexity of different payment networks. These processors implement sophisticated event systems and webhooks to handle payment state changes, retry logic, and fraud detection. This programmatic approach to payment processing has made it possible for digital services to implement complex payment flows without managing direct relationships with financial institutions.

Digital payment platforms operate as self-contained systems with their own ledgers. While often presented as "digital wallets," these platforms effectively function as e-money institutions, subject to specific regulatory requirements. Users maintain balances within the platform, with transactions executing as internal ledger updates. This model has evolved furthest in Asian markets, where super-apps like WeChat Pay and Alipay process enormous transaction volumes within their ecosystems. These platforms must still interface with traditional banking systems for deposits and withdraws but can operate independently for transfers between users.

The cryptocurrency ecosystem has developed a parallel payment infrastructure. While blockchain networks provide the underlying transaction layer, practical implementation requires additional infrastructure. Cryptocurrency exchanges serve as key on and off-ramps, converting between traditional currency and digital assets. Wallet software manages key storage and transaction signing, while stablecoins facilitate faster settlement by avoiding traditional banking rails. However, this infrastructure still struggles with scalability and regulatory compliance, particularly around identity verification and anti-money laundering requirements.

\subsection{Limitations for AI Agents}
\label{subsec:payments_limitations}
Current payment infrastructure presents fundamental barriers to AI agents operating as autonomous economic actors. The most immediate challenges stem from identity and verification requirements designed around human actors. Payment systems require extensive Know Your Customer (KYC) and Anti-Money Laundering (AML) verification, including government-issued identification, physical addresses, and phone numbers. These systems often include manual review processes and CAPTCHA mechanisms specifically designed to prevent automated account creation and transactions. Even when verification is possible, it typically requires human intervention, breaking the potential for fully autonomous operation.

A crucial point is that modern payment infrastructure is deliberately designed to prevent automated participation. Payment processors like Stripe implement multiple layers of anti-automation measures - from browser fingerprinting to behavioral analysis - specifically to ensure human involvement in financial transactions. These are not only security features but core design principles of the system. Anti-bot measures are deeply embedded in every layer, from account creation through transaction processing, making the infrastructure inherently hostile to autonomous agent participation.

Technical and operational constraints further limit AI agent participation. Typical API rate limits of 100 requests per minute for standard payment endpoints would severely constrain agents operating at machine speed. Authorization flows frequently require human interaction through redirect flows or manual confirmation steps. Fraud prevention systems flag patterns common to AI agents as suspicious - such as rapid sequential transactions or operations from multiple IP addresses. Traditional session-based security models and human-oriented authentication methods create additional friction for automated systems. Settlement delays range from 24-48 hours for credit card payments to 3-5 days for international bank transfers, making them unsuitable for agents operating at machine speed.

Cross-border transactions face particularly severe restrictions. International payments often require additional verification steps, with some payment processors blocking transactions from certain regions entirely. Currency conversion adds both cost and complexity - typical forex fees range from 1-3\%, with additional spread costs during conversion. Cross-border payments must navigate complex regulatory frameworks including  Financial Action Task Force (FATF) requirements, local banking regulations, and regional data protection laws. These international constraints often result in higher decline rates, longer settlement times, and increased compliance requirements.

Business model limitations also restrict AI agent participation. Current fee structures are optimized for human-scale transactions, making micropayments economically unfeasible due to fixed transaction costs - typically \$0.30 plus a percentage for credit card transactions. Subscription models assume monthly billing cycles rather than dynamic usage patterns. Pricing models are typically fixed rather than programmable, preventing automated negotiation or real-time price adjustment. Usage-based billing systems often lack the granularity and flexibility needed for machine-speed service consumption.

Regulatory frameworks compound these challenges by assuming human actors in financial transactions. Compliance requirements expect human oversight and accountability, with audit trails designed for human review. Legal frameworks for financial liability and dispute resolution assume human decision-makers. While these regulations serve important consumer protection and security purposes, they create significant barriers to autonomous AI agent participation in financial transactions.

\subsection{Future Design Considerations}
\label{subsec:payments_future}
Future payment infrastructure must enable autonomous economic participation while maintaining integration with existing financial systems. This requires both protocol-level innovations and new operational frameworks. At the protocol level, payment systems need standardized interfaces for programmatic execution, condition verification, and real-time settlement. For example, payment APIs could expose endpoints for automated compliance checks and conditional transfers, enabling agents to verify and execute transactions without human intervention. These protocols would need to support atomic operations where complex multi-step transactions either complete entirely or roll back, preventing inconsistent states during agent interactions.

An example of this is the L402 protocol which demonstrates potential approaches to machine-friendly payments. L402 extends the original HTTP 402 status code to create a complete protocol for payment-required API access. When a server responds with L402, it includes machine-readable payment terms specifying the price, payment methods, and conditions for access. Clients can then complete payments through supported payment networks before retrying their request with proof of payment. This decoupling of payment negotiation from payment execution allows clients to optimize their payment strategy while giving servers flexibility in how they price and gate access. Such protocols could form the foundation for standardized payment interactions between AI agents and services.

Authentication and compliance infrastructure requires significant evolution. Rather than relying on human documentation, systems could implement cryptographic attestation protocols where agents prove their identity and authorization through verifiable credentials. Traditional KYC/AML processes could be augmented with continuous transaction monitoring specifically designed for agent behavior patterns. This would enable payment processors to maintain regulatory compliance while supporting autonomous operation. Such systems might implement risk scoring based on agent transaction history, delegation chains, and behavioral patterns rather than traditional credit metrics.

The payment infrastructure must adapt to support machine-scale operations. Payment processors could implement tiered fee structures optimized for high-frequency, low-value transactions. Rather than fixed fees that make micropayments impractical, dynamic pricing could adjust based on volume and market conditions. New protocols could enable real-time price discovery and automated negotiation between agents.

To support AI agents as economic participants, digital infrastructure needs enhancements that cater to their unique capabilities. Agents would benefit from digital wallets designed for autonomous operation, allowing them to manage funds and transact without human involvement.

This infrastructure evolution requires careful coordination between payment processors, financial institutions, and regulatory bodies. Each component must maintain backward compatibility while enabling new capabilities. The goal is to extend existing payment rails to support autonomous economic actors while preserving the security and reliability of current financial systems.

\section{Future Work}
\label{sec:future_work}
Enabling AI agents to participate fully in digital markets represents a transformation too vast for any single organization to accomplish. The infrastructure challenges outlined in this paper - from authentication to payments - are deeply interconnected and require coordinated evolution across multiple domains.

Rather than seeking perfect solutions immediately, the community should focus first on creating an end-to-end protocol stack that enables basic machine-to-machine service discovery, integration, and consumption. This initial version might rely on imperfect workarounds and existing infrastructure, but it would provide a crucial foundation for experimentation and learning. For instance, early implementations might leverage existing API gateways and payment processors while adding machine-readable service descriptions and basic automated authentication.

The long-term value will come from infrastructure specifically optimized for machine consumption. Each component of the stack presents unique challenges that merit deeper investigation:

Service discovery requires protocols capable of expressing service capabilities, requirements, and integration patterns to machines. This extends beyond current API documentation to include semantic descriptions of functionality, performance characteristics, and operational constraints. Initiatives like the llms.txt proposal demonstrate early steps in this direction, creating parallel machine-friendly layers alongside traditional human-centric websites. Both effective representation formats and discovery mechanisms that operate at machine scale are essential components of this evolution.

Authentication and authorization systems designed for ephemeral agents operating at machine speed are fundamental building blocks. This includes scalable identity verification without human intervention, reputation systems for short-lived entities, and delegation mechanisms that maintain security across complex agent hierarchies.

Interface protocols supporting dynamic capability negotiation and efficient machine-to-machine communication represent another critical area. While recent developments like Anthropic's computer use capability demonstrate the potential for agents to interact with existing human interfaces, purpose-built machine interfaces will probably prove more efficient and reliable. Standards for describing service capabilities, negotiating terms of service, and establishing trust between previously unknown parties are essential.

Payment infrastructure presents perhaps the most formidable barrier, as current systems are actively hostile to automated transactions. However, major players are beginning to address this challenge - both Coinbase and Stripe have developed agent toolkits enabling AI systems to transact on their networks. Future infrastructure must support diverse transaction patterns between autonomous agents, including high-frequency micropayments and complex multi-party settlements while maintaining security without human intervention.

Beyond individual components, understanding how these machine-speed markets will behave presents its own challenges. New economic models for agent interactions, monitoring systems capable of detecting anomalies at machine speed, and circuit breakers that maintain stability without unnecessarily constraining market operation are all critical elements.

Each of these areas represents a significant research and development challenge. Progress will require collaboration between academia, industry, and standards bodies to create infrastructure that is both technically sophisticated and practically deployable. While the path forward may be incremental, the potential value of enabling autonomous agent participation in digital markets makes this effort worthwhile.

\section{Conclusion}
The convergence of artificial intelligence and market economics represents a profound opportunity for advancing both fields.

The emergence of AI systems capable of perception, reasoning, and action - currently enabled by large language models but potentially by other architectures in the future - has created the foundation for artificial agents that can operate as autonomous participants in digital markets. These systems can perceive their environment, reason about opportunities, and take concrete actions through code generation and execution, mirroring how humans participate in economic systems.

While much attention focuses on developing increasingly sophisticated individual AI systems, enabling these agents to participate in markets could unlock even greater potential. Just as markets enable human intelligence to coordinate complex activities beyond any individual's capability, markets incorporating AI agents could dramatically enhance economic efficiency through continuous operation, perfect information sharing, and rapid adaptation to changing conditions.

Consider how current markets coordinate the construction of modern aircraft. Thousands of companies across the globe, each specializing in particular components, collaborate through market mechanisms to create machines of staggering complexity. No central authority dictates every detail, yet markets enable this massive coordination task. Now imagine this same dynamic operating at machine speed, with AI agents discovering and exploiting opportunities for value creation far faster than humans ever could.

While research teams have made significant progress simulating multi-agent systems in controlled environments, these remain theoretical exercises—like flight simulators instead of actual aviation. The difference between agents playing economic games and agents participating in real markets is the difference between understanding principles and transforming society. The infrastructure components examined in this paper—authentication systems, payment networks, service discovery mechanisms, and standardized interfaces—represent the crucial bridges between simulation and reality.

This vision of AI development through market participation offers a multiplier effect to the pursuit of artificial superintelligence through purely algorithmic means. We might enable new forms of intelligence to emerge from the interactions of millions of specialized agents, just as markets already demonstrate intelligence beyond their individual participants. However, this transformation must proceed thoughtfully, with deep consideration of human interests and societal impact. The goal is not to replace human economic activity but to create new forms of partnership that benefit society while carefully managing potential risks.

The path forward requires careful development of appropriate infrastructure while maintaining security and stability. Yet the potential benefits—more efficient resource allocation, faster innovation, and new forms of value creation—make this effort worthwhile. We have the core AI capabilities needed for market participation in systems that can perceive, reason, and act. Building the infrastructure to enable their participation represents our next great challenge in advancing both artificial intelligence and economic systems.

\section*{Acknowledgments}

We would like to thank Javier Álvarez Cid-Fuentes for his valuable feedback and contributions in reviewing this paper and Rosa Maria Badia for endorsing our Arxiv submission.

We would also like to thank ChatGPT and Claude Sonnet for providing feedback through the perspectives of Paul Graham, Elon Musk, Jeremy Howard and Marvin Minsky during the writing process.

\bibliographystyle{plainnat}
\bibliography{references}
\end{document}

%% file: figures/service_discovery_mindmap.tex
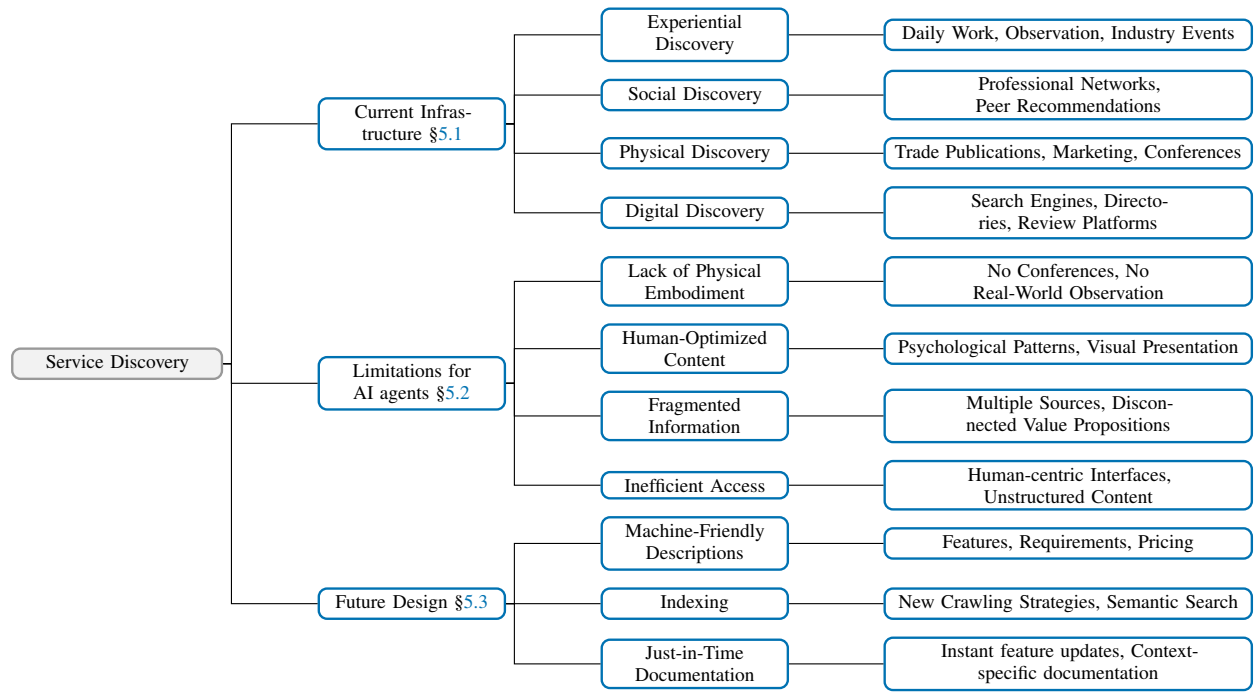
\begin{figure*}[!ht]
    \scriptsize
    \begin{adjustbox}{width=\textwidth}
        \begin{forest}
            mindmap
            [Service Discovery, fill=gray!45, parent
                [Current Infrastructure \S\ref{subsec:discovery_current}, node
                    [Experiential Discovery, node
                        [{Daily Work, Observation, Industry Events}, leaf]
                    ]
                    [Social Discovery, node
                        [{Professional Networks, Peer Recommendations}, leaf]
                    ]
                    [Physical Discovery, node
                        [{Trade Publications, Marketing, Conferences}, leaf]
                    ]
                    [Digital Discovery, node
                        [{Search Engines, Directories, Review Platforms}, leaf]
                    ]
                ]
                [Limitations for AI agents \S\ref{subsec:discovery_limitations}, node
                    [Lack of Physical Embodiment, node
                        [{No Conferences, No Real-World Observation}, leaf]
                    ]
                    [Human-Optimized Content, node
                        [{Psychological Patterns, Visual Presentation}, leaf]
                    ]
                    [Fragmented Information, node
                        [{Multiple Sources, Disconnected Value Propositions}, leaf]
                    ]
                    [Inefficient Access, node
                        [{Human-centric Interfaces, Unstructured Content}, leaf]
                    ]
                ]
                [Future Design \S\ref{subsec:discovery_future}, node
                    [Machine-Friendly Descriptions, node
                        [{Features, Requirements, Pricing}, leaf]
                    ]
                    [Indexing, node
                        [{New Crawling Strategies, Semantic Search}, leaf]
                    ]
                    [Just-in-Time Documentation, node
                        [{Instant feature updates, Context-specific documentation}, leaf]
                    ]
                ]
            ]   
        \end{forest}
    \end{adjustbox}
    \caption{Service discovery components and challenges.}
    \label{fig:service_discovery_mindmap}
\end{figure*} 

%% file: figures/identity_mindmap.tex
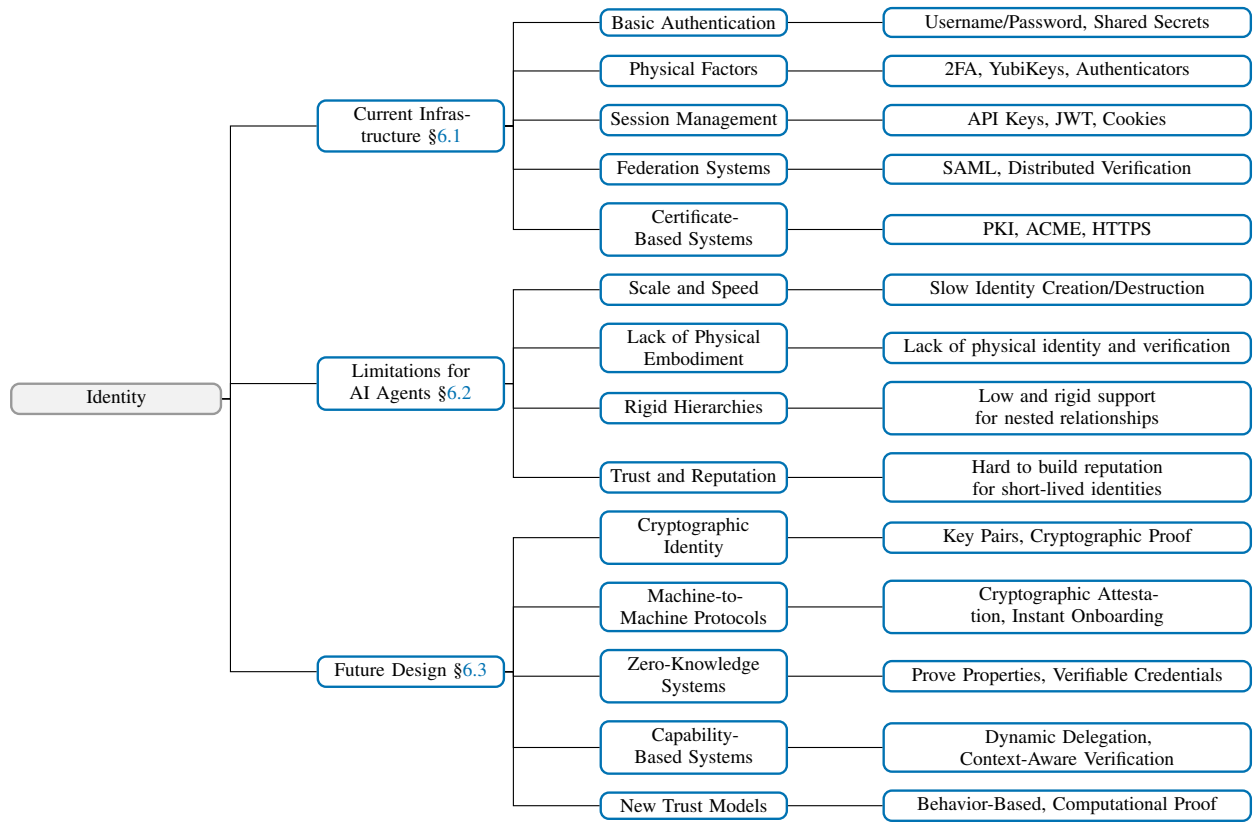
\begin{figure*}[!ht]
\scriptsize
    \begin{adjustbox}{width=\textwidth}
        \begin{forest}
        mindmap
        [Identity, fill=gray!45, parent
            [Current Infrastructure \S\ref{subsec:identity_current}, node
                [Basic Authentication, node
                    [{Username/Password, Shared Secrets}, leaf]
                ]
                [Physical Factors, node
                    [{2FA, YubiKeys, Authenticators}, leaf]
                ]
                [Session Management, node
                    [{API Keys, JWT, Cookies}, leaf]
                ]
                [Federation Systems, node
                    [{SAML, Distributed Verification}, leaf]
                ]
                [Certificate-Based Systems, node
                    [{PKI, ACME, HTTPS}, leaf]
                ]
            ]
            [Limitations for AI Agents \S\ref{subsec:identity_limitations}, node
                [Scale and Speed, node
                    [{Slow Identity Creation/Destruction}, leaf]
                ]
                [Lack of Physical Embodiment, node
                    [{Lack of physical identity and verification}, leaf]
                ]
                [Rigid Hierarchies, node
                    [{Low and rigid support for nested relationships}, leaf]
                ]
                [Trust and Reputation, node
                    [{Hard to build reputation for short-lived identities}, leaf]
                ]
            ]
            [Future Design \S\ref{subsec:identity_future}, node
                [Cryptographic Identity, node
                    [{Key Pairs, Cryptographic Proof}, leaf]
                ]
                [Machine-to-Machine Protocols, node
                    [{Cryptographic Attestation, Instant Onboarding}, leaf]
                ]
                [Zero-Knowledge Systems, node
                    [{Prove Properties, Verifiable Credentials}, leaf]
                ]
                [Capability-Based Systems, node
                    [{Dynamic Delegation, Context-Aware Verification}, leaf]
                ]
                [New Trust Models, node
                    [{Behavior-Based, Computational Proof}, leaf]
                ]
            ]
        ]   
        \end{forest}
    \end{adjustbox}
    \caption{Identity infrastructure components and challenges.}
    \label{fig:identity_mindmap}
\end{figure*}

%% file: figures/authorization_mindmap.tex
\begin{figure*}[!ht]
\scriptsize
    \begin{adjustbox}{width=\textwidth}
        \begin{forest}
        mindmap
        [Authorization, fill=gray!45, parent
            [Current Infrastructure \S\ref{subsec:authorization_current}, node
                [RBAC, node
                    [{Groups, Admin/Editor/Viewer Roles}, leaf]
                ]
                [ReBAC, node
                    [{Social/Organizational Relationships, Document Access}, leaf]
                ]
                [Token Systems, node
                    [{Stateful Sessions with Policies}, leaf]
                ]
                [Cross-Service Authorization, node
                    [{OAuth Flows, Delegated Access}, leaf]
                ]
                [Permission Management, node
                    [{Coarse-Grained, Infrequent Updates}, leaf]
                ]
            ]
            [Limitations for AI Agents \S\ref{subsec:authorization_limitations}, node
                [Static Permissions, node
                    [{Rigid Permission Systems, Lack of Action Delegation}, leaf]
                ]
                [Cross-Service Challenges, node
                    [{Continuous Delegation, Browser-based Approvals}, leaf]
                ]
                [Token Scalability, node
                    [{Revocation Challenges}, leaf]
                ]
            ]
            [Future Design \S\ref{subsec:authorization_future}, node
                [ABAC, node
                    [{Macaroons, Biscuits, Delegation Chains}, leaf]
                ]
                [Real-Time Policy Engines, node
                    [{Dynamic Evaluation, Predictive Permissions}, leaf]
                ]
                [Edge Verification, node
                    [{Distributed Evaluation, Audit Trails}, leaf]
                ]
                [Security Safeguards, node
                    [{Privilege Control, Context Protection}, leaf]
                ]
            ]
        ]   
        \end{forest}
    \end{adjustbox}
    \caption{Authorization infrastructure components and challenges.}
    \label{fig:authorization_mindmap}
\end{figure*}
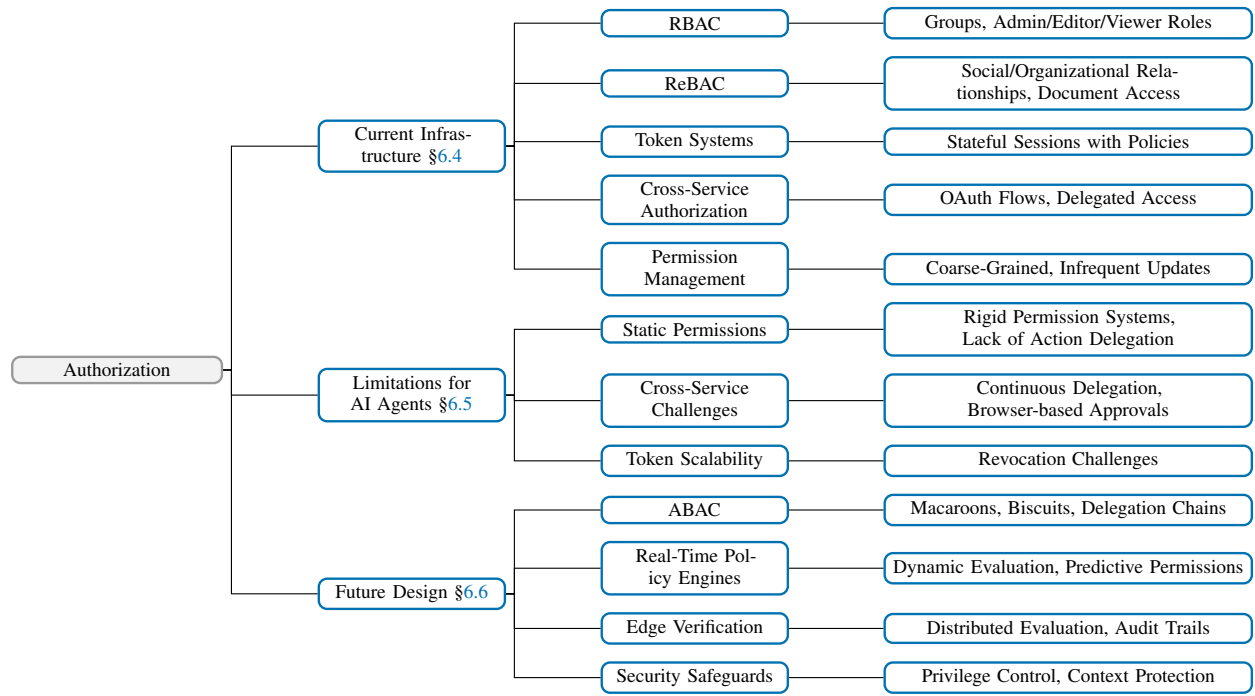 

%% file: figures/software_interfaces_mindmap.tex
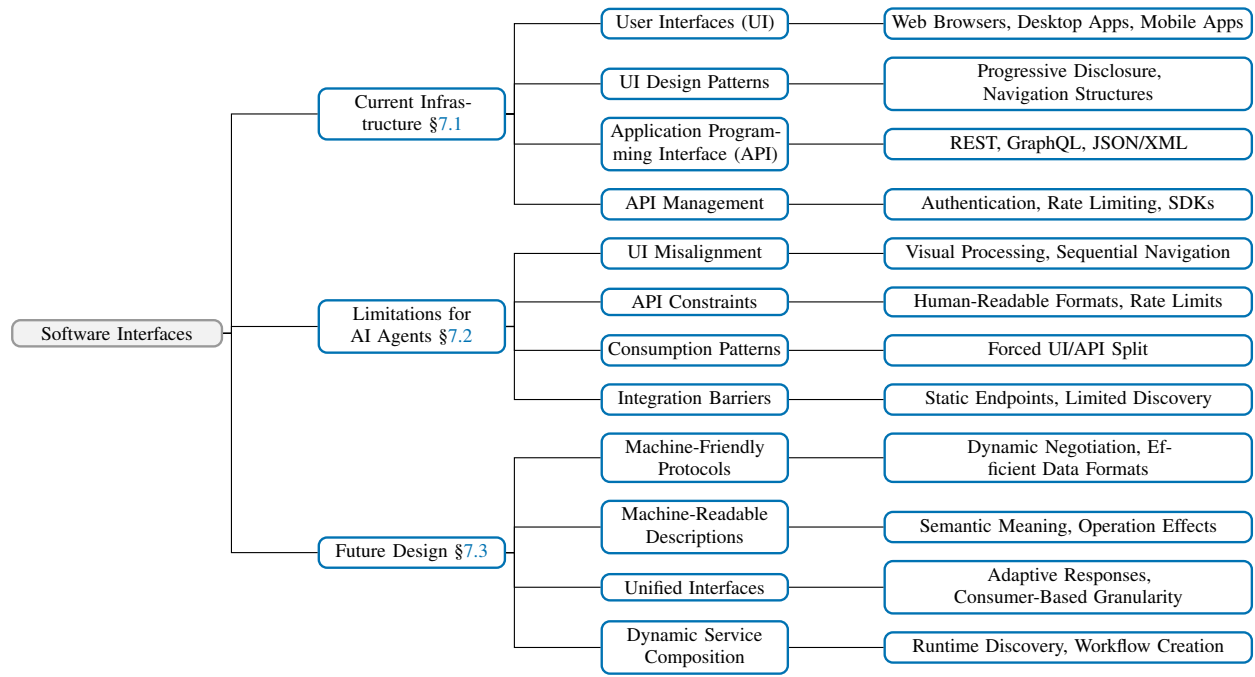
\begin{figure*}[!ht]
\scriptsize
    \begin{adjustbox}{width=\textwidth}
        \begin{forest}
        mindmap
        [Software Interfaces, fill=gray!45, parent
            [Current Infrastructure \S\ref{subsec:interfaces_current}, node
                [User Interfaces (UI), node
                    [{Web Browsers, Desktop Apps, Mobile Apps}, leaf]
                ]
                [UI Design Patterns, node
                    [{Progressive Disclosure, Navigation Structures}, leaf]
                ]
                [Application Programming Interface (API), node
                    [{REST, GraphQL, JSON/XML}, leaf]
                ]
                [API Management, node
                    [{Authentication, Rate Limiting, SDKs}, leaf]
                ]
            ]
            [Limitations for AI Agents \S\ref{subsec:interfaces_limitations}, node
                [UI Misalignment, node
                    [{Visual Processing, Sequential Navigation}, leaf]
                ]
                [API Constraints, node
                    [{Human-Readable Formats, Rate Limits}, leaf]
                ]
                [Consumption Patterns, node
                    [{Forced UI/API Split}, leaf]
                ]
                [Integration Barriers, node
                    [{Static Endpoints, Limited Discovery}, leaf]
                ]
            ]
            [Future Design \S\ref{subsec:interfaces_future}, node
                [Machine-Friendly Protocols, node
                    [{Dynamic Negotiation, Efficient Data Formats}, leaf]
                ]
                [Machine-Readable Descriptions, node
                    [{Semantic Meaning, Operation Effects}, leaf]
                ]
                [Unified Interfaces, node
                    [{Adaptive Responses, Consumer-Based Granularity}, leaf]
                ]
                [Dynamic Service Composition, node
                    [{Runtime Discovery, Workflow Creation}, leaf]
                ]
            ]
        ]   
        \end{forest}
    \end{adjustbox}
    \caption{Software interface components and challenges.}
    \label{fig:interfaces_mindmap}
\end{figure*}

%% file: figures/payments_mindmap.tex
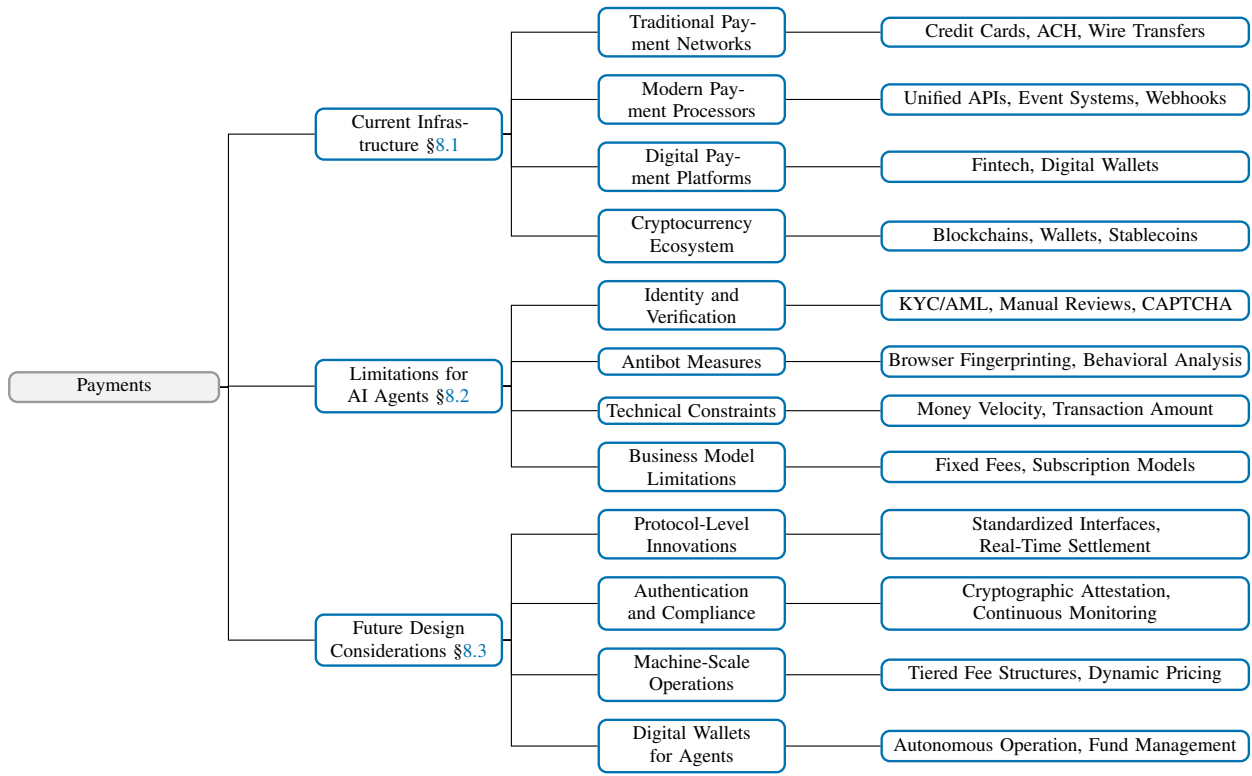
\begin{figure*}[!ht]
\scriptsize
    \begin{adjustbox}{width=\textwidth}
        \begin{forest}
        mindmap
        [Payments, fill=gray!45, parent
            [Current Infrastructure \S\ref{subsec:payments_current}, node
                [Traditional Payment Networks, node
                    [{Credit Cards, ACH, Wire Transfers}, leaf]
                ]
                [Modern Payment Processors, node
                    [{Unified APIs, Event Systems, Webhooks}, leaf]
                ]
                [Digital Payment Platforms, node
                    [{Fintech, Digital Wallets}, leaf]
                ]
                [Cryptocurrency Ecosystem, node
                    [{Blockchains, Wallets, Stablecoins}, leaf]
                ]
            ]
            [Limitations for AI Agents \S\ref{subsec:payments_limitations}, node
                [Identity and Verification, node
                    [{KYC/AML, Manual Reviews, CAPTCHA}, leaf]
                ]
                [Antibot Measures, node
                    [{Browser Fingerprinting, Behavioral Analysis}, leaf]
                ]
                [Technical Constraints, node
                    [{Money Velocity, Transaction Amount}, leaf]
                ]
                [Business Model Limitations, node
                    [{Fixed Fees, Subscription Models}, leaf]
                ]
            ]
            [Future Design Considerations \S\ref{subsec:payments_future}, node
                [Protocol-Level Innovations, node
                    [{Standardized Interfaces, Real-Time Settlement}, leaf]
                ]
                [Authentication and Compliance, node
                    [{Cryptographic Attestation, Continuous Monitoring}, leaf]
                ]
                [Machine-Scale Operations, node
                    [{Tiered Fee Structures, Dynamic Pricing}, leaf]
                ]
                [Digital Wallets for Agents, node
                    [{Autonomous Operation, Fund Management}, leaf]
                ]
            ]
        ]   
        \end{forest}
    \end{adjustbox}
    \caption{Payment infrastructure components and challenges.}
    \label{fig:payments_mindmap}
\end{figure*}

%% file: beyond_the_sum.bbl
\begin{thebibliography}{50}
\providecommand{\natexlab}[1]{#1}
\providecommand{\url}[1]{\texttt{#1}}
\expandafter\ifx\csname urlstyle\endcsname\relax
  \providecommand{\doi}[1]{doi: #1}\else
  \providecommand{\doi}{doi: \begingroup \urlstyle{rm}\Url}\fi

\bibitem[Ahuja et~al.(2019)Ahuja, Ma, Morency, and Sheikh]{Ahuja2019}
Chaitanya Ahuja, Shugao Ma, Louis-Philippe Morency, and Yaser Sheikh.
\newblock To react or not to react: End-to-end visual pose forecasting for
  personalized avatar during dyadic conversations.
\newblock 2019.
\newblock URL \url{https://arxiv.org/abs/1910.02181v1}.

\bibitem[Arthur(2009)]{arthur2009nature}
W.~Brian Arthur.
\newblock \emph{The Nature of Technology: What It Is and How It Evolves}.
\newblock Free Press, New York, USA, 2009.

\bibitem[Baltrušaitis et~al.(2017)Baltrušaitis, Ahuja, and
  Morency]{Baltrušaitis2017}
Tadas Baltrušaitis, Chaitanya Ahuja, and Louis-Philippe Morency.
\newblock "multimodal machine learning: A survey and taxonomy".
\newblock 2017.
\newblock URL \url{https://arxiv.org/abs/1705.09406v2}.

\bibitem[Bengio and Frasconi(1995)]{Bengio1995}
Y.~Bengio and P.~Frasconi.
\newblock Diffusion of context and credit information in markovian models.
\newblock 1995.
\newblock URL \url{https://arxiv.org/abs/9510101v1}.

\bibitem[Bergstra et~al.(2012)Bergstra, Yamins, and Cox]{Bergstra2012}
J.~Bergstra, D.~Yamins, and D.~D. Cox.
\newblock Making a science of model search.
\newblock 2012.
\newblock URL \url{https://arxiv.org/abs/1209.5111v1}.

\bibitem[Bostrom(2014)]{bostrom2014superintelligence}
Nick Bostrom.
\newblock \emph{Superintelligence: Paths, Dangers, Strategies}.
\newblock Oxford University Press, Oxford, UK, 2014.

\bibitem[Botvinick et~al.(2017)Botvinick, Barrett, Battaglia, de~Freitas,
  Kumaran, Leibo, Lillicrap, Modayil, Mohamed, Rabinowitz, Rezende, Santoro,
  Schaul, Summerfield, Wayne, Weber, Wierstra, Legg, and
  Hassabis]{Botvinick2017}
M.~Botvinick, D.~G.~T. Barrett, P.~Battaglia, N.~de~Freitas, D.~Kumaran, J.~Z
  Leibo, T.~Lillicrap, J.~Modayil, S.~Mohamed, N.~C. Rabinowitz, D.~J. Rezende,
  A.~Santoro, T.~Schaul, C.~Summerfield, G.~Wayne, T.~Weber, D.~Wierstra,
  S.~Legg, and D.~Hassabis.
\newblock Building machines that learn and think for themselves: Commentary on
  lake et al., behavioral and brain sciences, 2017.
\newblock 2017.
\newblock URL \url{https://arxiv.org/abs/1711.08378v1}.

\bibitem[Brown et~al.(2020)Brown, Mann, Ryder, Subbiah, Kaplan, Dhariwal,
  Neelakantan, Shyam, Sastry, Askell, Agarwal, Herbert-Voss, Krueger, Henighan,
  Child, Ramesh, Ziegler, Wu, Winter, Hesse, Chen, Sigler, Litwin, Gray, Chess,
  Clark, Berner, McCandlish, Radford, Sutskever, and Amodei]{Brown2020}
Tom~B. Brown, Benjamin Mann, Nick Ryder, Melanie Subbiah, Jared Kaplan,
  Prafulla Dhariwal, Arvind Neelakantan, Pranav Shyam, Girish Sastry, Amanda
  Askell, Sandhini Agarwal, Ariel Herbert-Voss, Gretchen Krueger, Tom Henighan,
  Rewon Child, Aditya Ramesh, Daniel~M. Ziegler, Jeffrey Wu, Clemens Winter,
  Christopher Hesse, Mark Chen, Eric Sigler, Mateusz Litwin, Scott Gray,
  Benjamin Chess, Jack Clark, Christopher Berner, Sam McCandlish, Alec Radford,
  Ilya Sutskever, and Dario Amodei.
\newblock Language models are few-shot learners.
\newblock 2020.
\newblock URL \url{https://arxiv.org/abs/2005.14165}.

\bibitem[Brynjolfsson and McAfee(2011)]{brynjolfsson2011race}
Erik Brynjolfsson and Andrew McAfee.
\newblock \emph{Race against the machine: How the digital revolution is
  accelerating innovation, driving productivity, and irreversibly transforming
  employment and the economy}.
\newblock Brynjolfsson and McAfee, 2011.

\bibitem[Brynjolfsson and McAfee(2014)]{brynjolfsson2014second}
Erik Brynjolfsson and Andrew McAfee.
\newblock \emph{The Second Machine Age: Work, Progress, and Prosperity in a
  Time of Brilliant Technologies}.
\newblock W. W. Norton \& Company, 2014.

\bibitem[Cameron(2005)]{cameron2005laws}
Kim Cameron.
\newblock The laws of identity.
\newblock \emph{Microsoft Corp}, 12:\penalty0 8--11, 2005.

\bibitem[Clark and Storkey(2014)]{Clark2014}
Christopher Clark and Amos Storkey.
\newblock "mastering the game of go with deep neural networks and tree search".
\newblock 2014.
\newblock URL \url{https://arxiv.org/abs/1412.3409v2}.

\bibitem[Claudette(2024)]{claudette2024toolloop}
Claudette.
\newblock Tool loop – claudette, 2024.
\newblock URL \url{https://claudette.answer.ai/toolloop.html}.
\newblock Accessed: 2024-06-17.

\bibitem[{Codeium}(2024)]{windsurf}
{Codeium}.
\newblock {Windsurf: Codeium's AI-Powered Code Generation}, 2024.
\newblock URL \url{https://codeium.com/windsurf}.
\newblock Accessed: 2024-06-17.

\bibitem[{Cursor AI}(2024)]{cursor}
{Cursor AI}.
\newblock {Cursor: The AI-Powered Code Editor}, 2024.
\newblock URL \url{https://www.cursor.com/}.
\newblock Accessed: 2024-06-17.

\bibitem[Dix(2004)]{dix2004human}
Alan Dix.
\newblock \emph{Human-computer interaction}, volume~1.
\newblock Pearson Education, 2004.

\bibitem[Douceur(2002)]{douceur2002sybil}
John~R Douceur.
\newblock The sybil attack.
\newblock In \emph{International workshop on peer-to-peer systems}, pages
  251--260. Springer, 2002.

\bibitem[GitHub(2024)]{github2024survey}
GitHub.
\newblock Survey: The ai wave continues to grow on software development teams -
  the github blog, 2024.
\newblock URL
  \url{https://github.blog/news-insights/research/survey-ai-wave-grows/}.
\newblock Accessed: 2024-06-17.

\bibitem[{GitHub, Inc.}(2024)]{copilot}
{GitHub, Inc.}
\newblock {GitHub Copilot: Your AI Pair Programmer}, 2024.
\newblock URL \url{https://github.com/features/copilot}.
\newblock Accessed: 2024-06-17.

\bibitem[Goodfellow et~al.(2013)Goodfellow, Erhan, Carrier, Courville, Mirza,
  Hamner, Cukierski, Tang, Thaler, Lee, Zhou, Ramaiah, Feng, Li, Wang,
  Athanasakis, Shawe-Taylor, Milakov, Park, Ionescu, Popescu, Grozea, Bergstra,
  Xie, Romaszko, Xu, Chuang, and Bengio]{Goodfellow2013}
Ian~J. Goodfellow, Dumitru Erhan, Pierre~Luc Carrier, Aaron Courville, Mehdi
  Mirza, Ben Hamner, Will Cukierski, Yichuan Tang, David Thaler, Dong-Hyun Lee,
  Yingbo Zhou, Chetan Ramaiah, Fangxiang Feng, Ruifan Li, Xiaojie Wang,
  Dimitris Athanasakis, John Shawe-Taylor, Maxim Milakov, John Park, Radu
  Ionescu, Marius Popescu, Cristian Grozea, James Bergstra, Jingjing Xie,
  Lukasz Romaszko, Bing Xu, Zhang Chuang, and Yoshua Bengio.
\newblock Challenges in representation learning: A report on three machine
  learning contests.
\newblock 2013.
\newblock URL \url{https://arxiv.org/abs/1307.0414}.

\bibitem[Greenstein(2019)]{greenstein2019digital}
Shane Greenstein.
\newblock Digital infrastructure.
\newblock \emph{Economics of infrastructure investment () University of Chicago
  Press}, 2019.

\bibitem[Haldane et~al.(2008)Haldane, Millard, and Saporta]{haldane2008future}
Andrew~G Haldane, Stephen Millard, and Victoria Saporta.
\newblock \emph{The future of payment systems}.
\newblock Routledge, 2008.

\bibitem[Hayek(1945)]{hayek1945knowledge}
F.~A. Hayek.
\newblock The use of knowledge in society.
\newblock \emph{The American Economic Review}, 35\penalty0 (4):\penalty0
  519--530, 1945.

\bibitem[Kanervisto et~al.(2020{\natexlab{a}})Kanervisto, Karttunen, and
  Hautamäki]{Kanervisto2020a}
Anssi Kanervisto, Janne Karttunen, and Ville Hautamäki.
\newblock "playing to learn: A study of minecraft".
\newblock 2020{\natexlab{a}}.
\newblock URL \url{https://arxiv.org/abs/2005.03374v1}.

\bibitem[Kanervisto et~al.(2020{\natexlab{b}})Kanervisto, Kinnunen, and
  Hautamäki]{Kanervisto2020b}
Anssi Kanervisto, Tomi Kinnunen, and Ville Hautamäki.
\newblock General characterization of agents by states they visit.
\newblock 2020{\natexlab{b}}.
\newblock URL \url{https://arxiv.org/abs/2012.01244v3}.

\bibitem[Kanervisto et~al.(2020{\natexlab{c}})Kanervisto, Scheller, and
  Hautamäki]{Kanervisto2020c}
Anssi Kanervisto, Christian Scheller, and Ville Hautamäki.
\newblock Action space shaping in deep reinforcement learning.
\newblock 2020{\natexlab{c}}.
\newblock URL \url{https://arxiv.org/abs/2004.00980v2}.

\bibitem[Kaplan et~al.(2020{\natexlab{a}})Kaplan, McCandlish, Henighan, Brown,
  Chess, Child, Gray, Radford, Wu, and Amodei]{Kaplan2020}
Jared Kaplan, Sam McCandlish, Tom Henighan, Tom~B. Brown, Benjamin Chess, Rewon
  Child, Scott Gray, Alec Radford, Jeffrey Wu, and Dario Amodei.
\newblock Scaling laws for neural language models.
\newblock 2020{\natexlab{a}}.
\newblock URL \url{https://arxiv.org/abs/2001.08361}.

\bibitem[Kaplan et~al.(2020{\natexlab{b}})Kaplan, McCandlish, Henighan, Brown,
  Chess, Child, Gray, Radford, Wu, and Amodei]{kaplan2020scaling}
Jared Kaplan, Sam McCandlish, Tom Henighan, Tom~B. Brown, Benjamin Chess, Rewon
  Child, Scott Gray, Alec Radford, Jeffrey Wu, and Dario Amodei.
\newblock Scaling laws for neural language models.
\newblock 2020{\natexlab{b}}.
\newblock URL \url{https://arxiv.org/abs/2001.08361}.

\bibitem[Karttunen et~al.(2019)Karttunen, Kanervisto, Kyrki, and
  Hautamäki]{Karttunen2019}
Janne Karttunen, Anssi Kanervisto, Ville Kyrki, and Ville Hautamäki.
\newblock From video game to real robot: The transfer between action spaces.
\newblock 2019.
\newblock URL \url{https://arxiv.org/abs/1905.00741v2}.

\bibitem[Lee et~al.(2022)Lee, Ahuja, Liang, Natu, and Morency]{Lee2022}
Dong~Won Lee, Chaitanya Ahuja, Paul~Pu Liang, Sanika Natu, and Louis-Philippe
  Morency.
\newblock Multimodal lecture presentations dataset: Understanding multimodality
  in educational slides.
\newblock 2022.
\newblock URL \url{https://arxiv.org/abs/2208.08080v1}.

\bibitem[Li et~al.(2022)Li, Lu, Guo, Duan, Jannu, Jenks, Majumder, Green,
  Svyatkovskiy, Fu, and Sundaresan]{li2022automatingcodereviewactivities}
Zhiyu Li, Shuai Lu, Daya Guo, Nan Duan, Shailesh Jannu, Grant Jenks, Deep
  Majumder, Jared Green, Alexey Svyatkovskiy, Shengyu Fu, and Neel Sundaresan.
\newblock Automating code review activities by large-scale pre-training, 2022.
\newblock URL \url{https://arxiv.org/abs/2203.09095}.

\bibitem[Lopez~de Prado(2018)]{lopez2018financial}
Marcos Lopez~de Prado.
\newblock \emph{Advances in Financial Machine Learning}.
\newblock Wiley, 2018.

\bibitem[Malone and Bernstein(2022)]{malone2022handbook}
Thomas~W Malone and Michael~S Bernstein.
\newblock \emph{Handbook of collective intelligence}.
\newblock MIT press, 2022.

\bibitem[Marcus(2018)]{Marcus2018}
Gary Marcus.
\newblock Deep learning: A critical appraisal.
\newblock 2018.
\newblock URL \url{https://arxiv.org/abs/1801.00631}.

\bibitem[Marinescu(2022)]{marinescu2022cloud}
Dan~C Marinescu.
\newblock \emph{Cloud computing: theory and practice}.
\newblock Morgan Kaufmann, 2022.

\bibitem[Milani et~al.(2023)Milani, Kanervisto, Ramanauskas, Schulhoff,
  Houghton, and Shah]{Milani2023}
Stephanie Milani, Anssi Kanervisto, Karolis Ramanauskas, Sander Schulhoff,
  Brandon Houghton, and Rohin Shah.
\newblock Bedd: The minerl basalt evaluation and demonstrations dataset for
  training and benchmarking agents that solve fuzzy tasks.
\newblock 2023.
\newblock URL \url{https://arxiv.org/abs/2312.02405v1}.

\bibitem[Mnih et~al.(2013)Mnih, Kavukcuoglu, Silver, Graves, Antonoglou,
  Wierstra, and Riedmiller]{Mnih2013}
Volodymyr Mnih, Koray Kavukcuoglu, David Silver, Alex Graves, Ioannis
  Antonoglou, Daan Wierstra, and Martin Riedmiller.
\newblock "playing atari with deep reinforcement learning".
\newblock 2013.
\newblock URL \url{https://arxiv.org/abs/1312.5602v1}.

\bibitem[Nisan et~al.(2007)Nisan, Roughgarden, Tardos, and
  Vazirani]{nisan2007algorithmic}
Noam Nisan, Tim Roughgarden, Eva Tardos, and Vijay~V Vazirani, editors.
\newblock \emph{Algorithmic game theory}.
\newblock Cambridge University Press, 2007.

\bibitem[Nyatsanga et~al.(2023)Nyatsanga, Kucherenko, Ahuja, Henter, and
  Neff]{Nyatsanga2023}
Simbarashe Nyatsanga, Taras Kucherenko, Chaitanya Ahuja, Gustav~Eje Henter, and
  Michael Neff.
\newblock A comprehensive review of data-driven co-speech gesture generation.
\newblock 2023.
\newblock URL \url{https://arxiv.org/abs/2301.05339v4}.

\bibitem[Replit(2024)]{replit2024agent}
Replit.
\newblock Replit agent | replit docs, 2024.
\newblock URL \url{https://docs.replit.com/replitai/agent}.
\newblock Accessed: 2024-06-17.

\bibitem[Schelling(1978)]{schelling1978micromotives}
Thomas~C. Schelling.
\newblock \emph{Micromotives and Macrobehavior}.
\newblock W. W. Norton \& Company, New York, USA, 1978.

\bibitem[Schumpeter(1942)]{schumpeter1942capitalism}
Joseph~A. Schumpeter.
\newblock \emph{Capitalism, Socialism, and Democracy}.
\newblock Harper \& Brothers, New York, USA, 1942.

\bibitem[Silver et~al.(2017)Silver, Hubert, Schrittwieser, Antonoglou, Lai,
  Guez, Lanctot, Sifre, Kumaran, Graepel, Lillicrap, Simonyan, and
  Hassabis]{Silver2017}
David Silver, Thomas Hubert, Julian Schrittwieser, Ioannis Antonoglou, Matthew
  Lai, Arthur Guez, Marc Lanctot, Laurent Sifre, Dharshan Kumaran, Thore
  Graepel, Timothy Lillicrap, Karen Simonyan, and Demis Hassabis.
\newblock "mastering chess and shogi by self-play with a general reinforcement
  learning algorithm".
\newblock 2017.
\newblock URL \url{https://arxiv.org/abs/1712.01815v1}.

\bibitem[Singh and Pal(2014)]{singh2014survey}
Ved~Prakash Singh and Preet Pal.
\newblock Survey of different types of captcha.
\newblock \emph{International Journal of Computer Science and Information
  Technologies}, 5\penalty0 (2):\penalty0 2242--2245, 2014.

\bibitem[Smith(1776)]{smith1776wealth}
Adam Smith.
\newblock \emph{An Inquiry into the Nature and Causes of the Wealth of
  Nations}.
\newblock W. Strahan and T. Cadell, London, UK, 1776.

\bibitem[Vaswani et~al.(2017)Vaswani, Shazeer, Parmar, Uszkoreit, Jones, Gomez,
  Kaiser, and Polosukhin]{Vaswani2017}
Ashish Vaswani, Noam Shazeer, Niki Parmar, Jakob Uszkoreit, Llion Jones,
  Aidan~N. Gomez, Lukasz Kaiser, and Illia Polosukhin.
\newblock Attention is all you need.
\newblock 2017.
\newblock URL \url{https://arxiv.org/abs/1706.03762}.

\bibitem[Wang et~al.(2023)Wang, Xie, Jiang, Mandlekar, Xiao, Zhu, Fan, and
  Anandkumar]{voyager}
Guanzhi Wang, Yuqi Xie, Yunfan Jiang, Ajay Mandlekar, Chaowei Xiao, Yuke Zhu,
  Linxi Fan, and Anima Anandkumar.
\newblock Voyager: An open-ended embodied agent with large language models,
  2023.
\newblock URL \url{https:%arxiv.org/abs/2305.16291}.

\bibitem[Xi et~al.(2023)Xi, Chen, Guo, He, Ding, Hong, Zhang, Wang, Jin, Zhou,
  Zheng, Fan, Wang, Xiong, Zhou, Wang, Jiang, Zou, Liu, Yin, Dou, Weng, Cheng,
  Zhang, Qin, Zheng, Qiu, Huang, and Gui]{ai_agents_survey}
Zhiheng Xi, Wenxiang Chen, Xin Guo, Wei He, Yiwen Ding, Boyang Hong, Ming
  Zhang, Junzhe Wang, Senjie Jin, Enyu Zhou, Rui Zheng, Xiaoran Fan, Xiao Wang,
  Limao Xiong, Yuhao Zhou, Weiran Wang, Changhao Jiang, Yicheng Zou, Xiangyang
  Liu, Zhangyue Yin, Shihan Dou, Rongxiang Weng, Wensen Cheng, Qi~Zhang,
  Wenjuan Qin, Yongyan Zheng, Xipeng Qiu, Xuanjing Huang, and Tao Gui.
\newblock The rise and potential of large language model based agents: A
  survey, 2023.
\newblock URL \url{https:%arxiv.org/abs/2309.07864}.

\bibitem[Zhu et~al.(2023)Zhu, Chen, Tian, Tao, Su, Yang, Huang, Li, Lu, Wang,
  Qiao, Zhang, and Dai]{zhu2023ghost}
Xizhou Zhu, Yuntao Chen, Hao Tian, Chenxin Tao, Weijie Su, Chenyu Yang, Gao
  Huang, Bin Li, Lewei Lu, Xiaogang Wang, Yu~Qiao, Zhaoxiang Zhang, and Jifeng
  Dai.
\newblock Ghost in the minecraft: Generally capable agents for open-world
  environments via large language models with text-based knowledge and memory,
  2023.
\newblock Shows how LLM-based agents overcome limitations of traditional RL
  approaches by dynamically generating actions instead of selecting from
  predefined ones. Traditional agents achieved only 30
  for LLM agents.

\bibitem[Ziegler et~al.(2019)Ziegler, Stiennon, Wu, Brown, Radford, Amodei,
  Christiano, and Irving]{Ziegler2019}
Daniel~M. Ziegler, Nisan Stiennon, Jeffrey Wu, Tom~B. Brown, Alec Radford,
  Dario Amodei, Paul Christiano, and Geoffrey Irving.
\newblock Fine-tuning language models from human preferences.
\newblock 2019.
\newblock URL \url{https://arxiv.org/abs/1909.08593v2}.

\end{thebibliography}
